\documentclass[
reprint,           
onecolumn,
superscriptaddress,
amsmath,           
amssymb,           
aps,               
prd,               
notitlepage,       
longbibliography,  
floatfix,          
nofootinbib,
]{revtex4-1}
\usepackage{amsmath}
\allowdisplaybreaks[4]
\usepackage{cancel}
\usepackage{extarrows}
\usepackage{tensor}     
\usepackage{float}
\usepackage[caption = false]{subfig} 
\usepackage[
colorlinks=true,      
citecolor=blue,         
linkcolor=blue,         
urlcolor=blue           
]{hyperref}             
\usepackage{bm}         %
\usepackage{xcolor}     %
\usepackage{lipsum}
\usepackage{color}      
\usepackage[utf8]{inputenc} 
\usepackage[section]{placeins} 
\usepackage{appendix}
\usepackage{units}
\usepackage{orcidlink}
\usepackage{tabularx}
\usepackage{adjustbox}
\usepackage{graphicx}
\newcommand{\nc}{\newcommand*}

\usepackage{float}

\usepackage{placeins}
\nc{\tr}{\rm{tr}}
\nc{\sig}{\sigma}
\nc{\om}{\omega}
\nc{\bt}{\beta}
\nc{\nb}{\nabla}
\nc{\eps}{\epsilon}
\nc{\epsS}{\epsilon_{\rm{S}}}
\nc{\epsV}{\epsilon_{\rm{V}}}
\nc{\epsT}{\epsilon_{\rm{T}}}
\nc{\osc}{\rm{osc}}
\nc{\DM}{\rm{TF}}
\nc{\FP}{\rm{FP}}
\nc{\eV}{\rm{eV}}
\nc{\figurewidth}{3.2in}
\nc{\xbar}{\bar{x}}
\nc{\rhoeq}{\rho_{\mathrm{eq}}}
\nc{\zeq}{z_{\mathrm{eq}}}
\nc{\tla}{\tilde{\lambda}}
\nc{\dt}{\delta}
\nc{\Dt}{\Delta}
\nc{\vj}{\hat{j}}
\nc{\vl}{\hat{l}}
\nc{\hx}{\hat{x}}
\nc{\hy}{\hat{y}}
\nc{\bj}{\bm{j}}
\nc{\mJ}{\mathcal{J}}
\nc{\mP}{\mathcal{P}}
\nc{\lbd}{\lambda}
\nc{\Msun}{M_\odot}
\nc{\app}{\approx}
\nc{\av}[1]{\langle #1 \rangle}
\nc{\eq}[1]{Eq.~\eqref{#1}}
\nc{\al}{\alpha}
\nc{\Xstar}{X_{\ast}}
\nc{\seq}{\sigma_{\mathrm{eq}}}
\nc{\fpbh}{f_{\mathrm{pbh}}}
\nc{\vth}{\hat{\theta}}
\nc{\vla}{\hat{\lambda}}
\nc{\vd}{\hat{d}}
\nc{\Mmin}{M_{\mathrm{min}}}
\nc{\rmd}{\mathrm{d}}
\nc{\mmin}{{m_{\mathrm{min}}}}
\nc{\mmax}{{m_{\mathrm{max}}}}
\nc{\mR}{\mathcal{R}}
\nc{\tmR}{\tilde{\mathcal{R}}}
\nc{\s}{\sigma}
\nc{\ogw}{\Omega_{\mathrm{GW}}}
\nc{\addref}{[\textcolor{red}{add ref}] }
\nc{\Om}{\Omega}
\nc{\gm}{\gamma}
\nc{\Gm}{\Gamma}
\nc{\kp}{\kappa}
\nc{\hbe}{\hat{\mathbf{e}}}
\nc{\gpcyr}{\mathrm{Gpc}^{-3}\,\mathrm{yr}^{-1}}
\nc{\Eq}[1]{Eq.~\eqref{#1}}
\nc{\Fig}[1]{Fig.~\ref{#1}}
\nc{\Table}[1]{Table~\ref{#1}}
\nc{\lvc}{LIGO/Virgo} 
\nc{\Sec}[1]{Sec.~\ref{#1}}
\nc{\eg}{\textit{e.g.~}}
\nc{\SNR}{\mathrm{SNR}}
\nc{\be}{\mathbf{\epsilon}}
\nc{\bn}{\mathbf{n}}
\nc{\bx}{\mathbf{x}}
\nc{\bk}{\mathbf{k}}
\nc{\bd}{\mathbf{d}}
\nc{\ba}{\mathbf{a}}
\nc{\bp}{\mathbf{p}}
\nc{\bnu}{\mathbf{\nu}}
\nc{\uni}{\mathrm{U}}
\nc{\logu}{\operatorname{\mathrm{log-U}}}
\nc{\RN}{\mathrm{RN}}
\nc{\BN}{\mathrm{BN}}
\nc{\GN}{\mathrm{GN}}
\nc{\mcN}{\mathcal{N}}
\nc{\GWB}{\mathrm{GW}}
\nc{\yr}{\mathrm{yr}}
\nc{\Am}{\mathcal{A}}
\nc{\Dm}{\mathcal{D}}
\nc{\Hm}{\mathcal{H}}
\nc{\sovast}{Soviet Ast.}
\nc{\hosc}{h_{\mathrm{osc}}}
\nc{\Posc}{\Psi_{\mathrm{osc}}}

\nc{\mrm}{\mathrm}
\nc{\BE}{B\scriptsize{AYES}\normalsize{E}\scriptsize{PHEM}\normalsize  }

\def\({\left(}
\def\){\right)}
\def\[{\left[}
\def\]{\right]}

\def\e{\begin{equation}}
\def\q{\end{equation}}
\def\m{\begin{eqnarray}}
\def\n{\end{eqnarray}}
\nc{\red}[1]{\textcolor{red}{#1}}
\begin{document}

\title{Spatial Correlation between Pulsars from Interfering Gravitational-Wave Sources in Massive Gravity}     

\author{Yu-Mei Wu\orcidlink{0000-0002-9247-5155}}
\email{ wuyumei@yzu.edu.cn} 
\affiliation{Center for Gravitation and Cosmology, College of Physical Science and Technology, Yangzhou University, Yangzhou, 225009, China}

\author{Yan-Chen Bi\orcidlink{0000-0002-9346-8715}}
\email{biyanchen@itp.ac.cn}
\affiliation{Institute of Theoretical Physics, Chinese Academy of Sciences,Beijing 100190, China}
\affiliation{School of Physical Sciences, 
    University of Chinese Academy of Sciences, 
    No. 19A Yuquan Road, Beijing 100049, China}
\author{Qing-Guo Huang\orcidlink{0000-0003-1584-345X}}
\email{huangqg@itp.ac.cn}
\affiliation{Institute of Theoretical Physics, Chinese Academy of Sciences,Beijing 100190, China}
\affiliation{School of Physical Sciences, 
    University of Chinese Academy of Sciences, 
    No. 19A Yuquan Road, Beijing 100049, China}
\affiliation{School of Fundamental Physics and Mathematical Sciences, Hangzhou Institute for Advanced Study, UCAS, Hangzhou 310024, China}


\begin{abstract}
In the nanohertz band, the spatial correlations in pulsar timing arrays (PTAs) produced by interfering gravitational waves (GWs) from multiple sources likely deviate from the traditional ones without interference under the assumption of an isotropic Gaussian ensemble. This work investigates the impact of such interference within the framework of massive gravity. Through simulations, we show that while the resulting correlation patterns can be described by Legendre expansions with coefficients that depend on the interference configuration, they remain predominantly quadrupolar ($l = 2$), with this feature becoming more pronounced as the graviton mass increases—reflecting both the tensorial polarizations and the modified GW dispersion. However, the interference introduces significant variability in the angular correlation, making it difficult to distinguish massive gravity from general relativity based on a single realization of the Universe. We conclude that beyond a fundamental constraint set by the PTA observation time, achieving a substantially tighter bound on the graviton mass is statistically challenging and observationally limited under realistic conditions.
\end{abstract}

	
\maketitle
	
	
\section{Introduction}
Whether the graviton has a non-zero mass remains one of the fundamental open questions in physics \citep{Fierz:1939ix}. Since the 1970s, significant theoretical efforts have been devoted to constructing a consistent and ghost-free theory of massive gravity \cite{vanDam:1970vg, Zakharov:1970cc,Vainshtein:1972sx,Boulware:1972yco}, leading to the development of prominent models such as the Dvali–Gabadadze–Porrati model \cite{Dvali:2000xg,Dvali:2000hr,Dvali:2000rv} and the de Rham–Gabadadze–Tolley model \cite{deRham:2010kj}. In parallel, extensive experimental and observational efforts have been made to detect or constrain the graviton mass—through, for example, tests of deviations from the Newtonian potential in the Solar System \cite{Bernus:2020szc}, observations of superradiant instabilities in supermassive black holes \cite{Brito:2013wya}, and measurements of weak gravitational lensing \cite{Choudhury:2002pu}. The advent of gravitational wave astronomy has opened a powerful new avenue for probing this question. Ground-based gravitational wave (GW) interferometers have placed bounds on the graviton mass using observed GW signals: the first detection, GW150914 \cite{LIGOScientific:2016lio} , yielded an upper limit of $m_g \le 1.2 \times 10^{-22}\mathrm{eV}/c^2$, which was subsequently improved to $m_g \le 1.27 \times 10^{-23}\mathrm{eV}/c^2$ using data from the Gravitational-Wave Transient Catalog (GWTC-3) \cite{LIGOScientific:2021sio}.

In addition to ground-based detectors, pulsar timing arrays (PTAs) have also recently made significant progress in detecting the stochastic gravitational-wave background (SGWB). PTAs are sensitive to nanohertz signals by monitoring the times of arrival of radio pulses emitted from millisecond pulsars \cite{1978SvA....22...36S,Detweiler:1979wn,1990ApJ...361..300F}, where such signals induce differences between the expected and actual arrival times—known as timing residuals. In particular, the SGWB—whether of astrophysical or cosmological origin \citep{Bi:2023tib,Wu:2023hsa}—generates timing residuals across different pulsars that exhibit a distinct spatial correlation \cite{Taylor:2015msb,Burke-Spolaor:2018bvk}.
After decades of observation, the North American Nanohertz Observatory for Gravitational Waves (NANOGrav, ~\citep{NANOGrav:2023hde,NANOGrav:2023gor}), the European PTA (EPTA) in collaboration with the Indian PTA (InPTA)~\citep{EPTA:2023sfo,EPTA:2023fyk}, the Parkes PTA (PPTA)~\citep{Zic:2023gta,Reardon:2023gzh}, and the Chinese PTA (CPTA)~\citep{Xu:2023wog} have found evidence for a stochastic signal consistent with the Hellings-Downs (HD) correlation \citep{Hellings:1983fr}, which is considered a smoking gun signature of the SGWB predicted by general relativity (GR).
However, modified gravity theories often predict different characteristic correlation curves~\cite{Bi:2023ewq, Chen:2023uiz,Wu:2021kmd,Chen:2021ncc}. Specifically, if the graviton possesses a nonvanishing mass, the dispersion relation of GWs would be modified, resulting in spatial correlations of timing residuals that deviate from the HD curve ~\citep{Lee:2010cg,Liang:2021bct,Wu:2023pbt}. These deviations open up the possibility of probing massive gravity using PTA data~\citep{Wu:2023rib}. Taking into account both this altered spatial correlation and the existence of a minimum frequency implied by a nonzero graviton mass, the NANOGrav 15-year data set places a tighter upper bound on the graviton mass than ground-based detectors, with $m_g \le 8.2 \times 10^{-24} \mathrm{eV}/c^2$~\citep{Wu:2023rib}.

Nevertheless, recent studies have highlighted that if the SGWB originates from supermassive black hole binaries, numerous sources in the nanohertz band emit gravitational waves at closely spaced frequencies, leading to interference effects~\citep{Roebber:2016jzl,Allen:2022dzg}. Consequently, the actual spatial correlation may differ from the HD curve even within the framework of GR~\citep{Allen:2022dzg,Romano:2023zhb,Bernardo:2022xzl,Bernardo:2023bqx}. While Ref.~\citep{Allen:2022dzg} interprets the HD curve as the ensemble average over an imaginary set of infinitely many independent Universes, where interference effects give rise to an irreducible variance—commonly referred to as the  ``cosmic variance''—in any single Universe, Ref.~\citep{Wu:2024xkp} argues that, in our uniquely realized Universe, the spatial correlation is definite and determined by the specific interference scenario. However, the lack of access to detailed source information, such as phases and locations, results in a loss of predictability for the true correlation pattern.
Ref.~\citep{Wu:2024xkp} further analyzes the features of spatial correlation curves under various interference configurations, suggesting the possibility that the diversity of interference patterns could lead to degeneracies between the correlation expected in GR and those predicted by modified gravity theories. More critically, an SGWB originating from alternative theories would likewise exhibit interference effects, making it seemingly even more challenging to distinguish between competing theoretical models based on the observed correlation patterns. However, previous studies~\citep{Bernardo:2023pwt,Bernardo:2023zna,Bernardo:2023jhs,Bernardo:2024bdc} have typically discussed correlations for modified gravity within a Gaussian ensemble from a statistical perspective. In particular, Ref.~\citep{Bernardo:2023pwt} elaborates the potential of using correlation measurements that are only limited by cosmic variance to test gravity, and expressed optimism about distinguishing GR from alternative theories using the first few multipoles.

In this work, we aim to investigate the spatial correlation patterns induced by realistic interference effects in the presence of a nonzero graviton mass, and to reassess the true discriminatory power of PTAs in testing massive gravity by simultaneously accounting for the correlation structure, its frequency-dependent behavior, and the presence of a cutoff frequency implied by the dispersion relation. We adopt natural units by setting the speed of light $c = 1$ and the reduced Planck constant $\hbar = 1$. The Earth is placed at the coordinate origin. Unit vectors are denoted by boldface symbols with hats, such as $\hat{\bm{\Omega}}$ and $\hat{\bm{p}}$. Latin subscripts are used with specific conventions throughout the paper: ``$a$'' and ``$b$'' denote abstract tensor indices; 
``$j$'' and ``$k$'' label GW sources; and ``A" and ``B" label pulsars.


\section{Signal Correlations from Interfering Sources}\label{sec2}
For GW propagating in the direction $\hat{\bm{\Om}}$, the four-wavevector in massive gravity is given by
\m
k^\mu = (\om, |\bm{k}| \hat{\bm{\Om}}) = 2\pi f (1, \eta \hat{\bm{\Om}}),
\n
where $\om=2\pi f$ is the angular frequency associated with the GW frequency $f$, $\bm{k}$ is the wave vector, and the factor $\eta$ characterizes the dispersion relation:
\m
\eta \equiv \frac{|\bm{k}|}{\om} = \sqrt{1 - \left( \frac{m_g}{2\pi f} \right)^2}.
\label{eta}
\n
Here, $m_g$ represents the graviton mass. By definition, $\eta$ is the inverse of the phase velocity $v_p=\om/|\bm{k}|$. In the specific case of massive gravity, the dispersion relation further leads to $\eta$ coinciding with the GW group velocity, namely $v_g\equiv d\om/d|\bm{k}|=\eta$, and the physical requirement that $v_g$ does not exceed the speed of light is fully consistent with the constraint $\eta \le 1$ imposed by \Eq{eta}. 

In massive gravity, there are theoretically five polarization modes: two helicity-2 (tensor) modes, two helicity-1 (vector) modes, and one helicity-0 (scalar) mode. However, in realistic astrophysical scenarios, vector modes are unlikely to be generated through natural processes~\cite{deRham:2014zqa} and have been shown to be disfavored \cite{Bernardo:2023mxc,Chen:2023uiz}, while scalar excitations are expected to be suppressed due to the Vainshtein screening mechanism \cite{deRham:2014zqa}. Therefore, in this work, we focus exclusively on the two tensorial polarization modes, as in GR.
Following \cite{Allen:2022dzg}, we assume that the GW sources are unpolarized, meaning they emit identical intrinsic amplitudes in both tensor polarization modes. Under this assumption, the waveform of the $j$-th source takes the form
\m
h_j^{+}(t,\bm{x})=A_j \cos[2\pi f_j (t-\eta 
\hat{\bm{\Om}}_j\cdot \bm{x})+\phi_j], \qquad 
h_j^{\times}(t,\bm{x})=A_j \sin[2\pi f_j (t-\eta \hat{\bm{\Om}}_j\cdot \bm{x})+\phi_j],
\n
where $A_j$ and $\phi_j$ represent the amplitude and phase of the GW source, respectively; so the total GW strain is given by
\m
h_{ab}(t,\bm{x})=\sum_j h_j^{+}(t,\bm{x})\varepsilon_{ab}^{+}(\hat{\bm{\Om}}_j)+h_j^{\times}(t,\bm{x})\varepsilon_{ab}^{\times}(\hat{\bm{\Om}}_j)
\n
with $\varepsilon_{ab}^{+}$ and $\varepsilon_{ab}^{\times}$ the normalized polarization tensors.
For convenience, we introduce the complex waveform $h_j(t,\bm{x})=h_j^{+}(t,\bm{x})+i \,h_j^{\times}(t,\bm{x})=A_j  e^{i[2\pi f_j (t-\eta \hat{\bm{\Om}}_j\cdot \bm{x})+\phi_j]}$, and adopt the circular polarization basis $\varepsilon_{ab}=\varepsilon_{ab}^{+}-i\, \varepsilon_{ab}^{\times}$ to rewrite $h_{ab}(t,\bm{x})$ as
\m
h_{ab}(t,\bm{x})&=&\mathcal{R} \sum_j h_j(t,\bm{x})\varepsilon_{ab}(\hat{\bm{\Om}}_j) \notag \\
&=&\frac{1}{2}\sum_j A_j \left\{ e^{i\[2\pi f_j (t-\eta \hat{\bm{\Om}}_j\cdot \bm{x})+\phi_j\]}\varepsilon_{ab}(\hat{\bm{\Om}}_j)+e^{-i\[2\pi f_j (t-\eta \hat{\bm{\Om}}_j\cdot \bm{x})+\phi_j\]}\varepsilon_{ab}^*(\hat{\bm{\Om}}_j)\right\}
\n

For an SGWB produced by a fixed set of GW sources, the radio pulse emitted by a pulsar located at position $L\hat{\bm{p}}$ experiences a redshift as it travels to the Earth. This redshift is given by
\m
Z(t)&=&\mathcal{R} \sum_j \Dt h_j(t, L\hat{\bm{p}}) F(\hat{\bm{\Om}}_j)\notag \\
&=&\frac{1}{2}\sum_j A_j\left\{e^{i(2\pi f_j t +\phi_j)}\[1-e^{-i2\pi f_j L(1+ \eta \hat{\bm{\Om}}_j \cdot \hat{\bm{p}})}\]F(\hat{\bm{\Om}}_j)+e^{-i(2\pi f_j t +\phi_j)}\[1-e^{i2\pi f_j L(1+ \eta \hat{\bm{\Om}}_j \cdot \hat{\bm{p}})}\]F^*(\hat{\bm{\Om}}_j)\right\}
\n 
Here, $\Delta h_j(t, L\hat{\bm{p}})$ denotes the complex GW strain difference between the Earth term, evaluated at time $t$, and the pulsar term, evaluated at the retarded time $t - L$,
\m
 \Dt h_j(t, L\hat{\bm{p}})&=&h_j(t,\bm{0})-h_j(t-L,L\hat{\bm{p}}),
\n
and $F(\hat{\bm{\Omega}}_j)$ represents the complex antenna response function,  which combines the standard ``plus” and ``cross” polarization components as
\m
F(\hat{\bm{\Om}}_j)&=&F^{+}(\hat{\bm{\Om}}_j)-i\, F^{\times}(\hat{\bm{\Om}}_j),
\n
with $F^{+,\times}(\hat{\bm{\Omega}}_j)$ related to the polarization tensor $\varepsilon_{ab}^{+,\times}$ via~\cite{Chamberlin:2011ev}
\e
F^{+,\times}(\hat{\bm{\Omega}}_j)=\frac{\hat{p}^{a}\hat{p}^{b}}{2(1+\eta \hat{\bm{\Omega}}_j\cdot \hat{\bm{p}})}\varepsilon_{ab}^{+,\times}.
\label{F+x}
\q

The redshift in the pulse frequency leads to an anomalous timing residual in the pulse time of arrival (ToA), given by the time integral of the redshift. PTAs identify the presence of an SGWB by exploiting its characteristic angular correlation pattern across pulsars, which distinguishes it from various noise sources in the ToAs. Since the redshift and its corresponding timing residual share the same geometric dependence, the spatial correlation can be derived from either quantity. In this work, we focus on the redshift. After a long time of observation $T$, typically spanning years to decades, the time-averaged redshift correlation between pulsars $A$ and $B$ is given by
\e
\begin{aligned}
\rho_{AB} &\equiv\overline{Z_{A}(t) Z_{B}(t)}\\
&= \sum_{j} \(c_j d_j^{*} + c_j^{*} d_j\) + \sum_{j\neq k}\[ c_j d_k^{*} \overline{\mathrm{e}^{2\pi i(f_j-f_k)t}} \mathrm{e}^{i(\phi_j-\phi_k)} + c_j^{*} d_k \overline{\mathrm{e}^{-2\pi i(f_j-f_k)t}} \mathrm{e}^{-i(\phi_j-\phi_k)}\] ,
\end{aligned}
\label{rhosimp1}
\q
with
\e
c_j = \frac{1}{2} A_j \[1 - \mathrm{e}^{-2\pi i f_j L_A (1 + \eta \hat{\bm{\Omega}}_j \cdot \hat{\bm{p}}_A)}\] F_A(\hat{\bm{\Omega}}_j), \qquad 
d_k = \frac{1}{2} A_k \[1 - \mathrm{e}^{-2\pi i f_k L_B (1 +\eta \hat{\bm{\Omega}}_k \cdot \hat{\bm{p}}_B)}\] F_B(\hat{\bm{\Omega}}_k)
\q
as the coefficients from pulsar $A$ and pulsar $B$ respectively. 
In \Eq{rhosimp1}, we decompose the redshift correlation into two parts: one arising from diagonal terms (i.e., identical source pairs) and the other from non-diagonal terms ($j \neq k$). While the diagonal contribution always survives, the contribution from non-diagonal pairs depends crucially on the frequency relationship between the sources. When the sources emit gravitational waves at significantly different frequencies relative to the observational time scale, the time-averaged term $\overline{\mathrm{e}^{2\pi i(f_j-f_k)t}}$ for $j \neq k$ vanishes. In contrast, if many sources emit at nearly the same frequency, i.e., $f_j \approx f_k$, the non-diagonal terms remain non-negligible. Although PTAs are sensitive to GWs over a broad frequency band spanning $10^{-9} \sim 10^{-7}$ Hz, data analysis is conducted in discrete Fourier frequency bins with a typical width of $\Delta f \sim 1/T \sim 1$ nHz. Each bin can contain on the order of $\mathcal{O}(10^3)$ sources with closely spaced frequencies. Therefore, within each frequency bin—where we effectively have $f_j = f_k$—the redshift correlation is expected to take the form
\e
\rho_{AB} = \sum_{j}\( c_j d_j^{*} + c_j^{*} d_j \)+ \sum_{j\neq k} \[c_j d_k^{*} \mathrm{e}^{i(\phi_j-\phi_k)} + c_j^{*} d_k \mathrm{e}^{-i(\phi_j-\phi_k)}\],
\q
where the non-diagonal contribution can be regarded as the result of ``interfering effect", which depends on the positions and  relative phases of the GW sources.

The correlation discussed above pertains to a single pulsar pair. In practice, however, PTAs typically monitor dozens of pulsars, resulting in multiple pulsar pairs that share similar angular separations. As a result, the quantity that is actually measured is a ``pulsar-averaged" correlation, which aggregates contributions from different pulsar pairs separated by the same angle $\gm= \cos^{-1}(\hat{\bm{p}}_A \cdot \hat{\bm{p}}_B)$,
\e
\begin{aligned}
\langle\rho_{AB}\rangle_p \!=\!& \sum_{j, k}\[ \langle c_j d_k^{*} \rangle_p \mathrm{e}^{i(\phi_j-\phi_k)} + \langle c_j^{*} d_k \rangle_p  \mathrm{e}^{-i(\phi_j-\phi_k)}\] \\
\!=\!&\frac{1}{4}\sum_{j,k} A_j A_k \mathrm{e}^{i(\phi_j-\phi_k)}\times\\
\!\!&\left\langle \[1\!-\! \mathrm{e}^{-2\pi i f_j L_A (1 + \eta \hat{\bm{\Omega}}_j \cdot \hat{\bm{p}}_A)}\!-\!\mathrm{e}^{2\pi i f_k L_B (1 + \eta \hat{\bm{\Omega}}_k \cdot \hat{\bm{p}}_B)}  \!+\! \mathrm{e}^{-2\pi i [f_j L_A (1 + \eta \hat{\bm{\Omega}}_j \cdot \hat{\bm{p}}_A)-f_k L_B (1 + \eta \hat{\bm{\Omega}}_k \cdot \hat{\bm{p}}_B)]} \]  F_A(\hat{\bm{\Omega}}_j)F_B^*(\hat{\bm{\Omega}}_k) \right\rangle_p \!+\! C.C.\\
\!=\!&\frac{1}{4}\sum_{j,k} A_j A_k \mathrm{e}^{i(\phi_j-\phi_k)} \langle F_A(\hat{\bm{\Omega}}_j)F_B^*(\hat{\bm{\Omega}}_k)\rangle_p+C.C..
\end{aligned}
\label{rhoave}
\q
Here, the subscript $p$ denotes an average over pulsars, and $C.C.$ indicates the complex conjugate of the preceding term. 
The exponential functions within the square brackets originate from the pulsar term. In the massless limit with $\eta=1$, these terms asymptotically oscillate to zero as $fL$ increases. For typical PTA scales with $fL \gtrsim 10^{2}$, it is therefore conventional to adopt the infinite-distance limit and neglect the pulsar term~\cite{Cornish:2013aba,Chamberlin:2011ev,Allen:2022dzg}. This approximation also remains valid in subluminal but relativistic cases, such as $v_g=\eta=1/2$~\cite{Bernardo:2022rif}. In contrast, for extremely subluminal, nonrelativistic scenarios, the finite pulsar distance effect becomes significant and could lead to substantial deviations in the correlation, particularly at small angular separations~\cite{Bernardo:2022rif,Domenech:2024pow}. In this work, we assume that nanohertz GWs propagate relativistically, and therefore neglect the pulsar term.


Before performing an explicit calculation of the redshift correlation $\langle\rho_{AB}\rangle_p$, we first analyze the geometric structure of the  two-point function $\langle F_A(\hat{\bm{\Omega}}_j)F_B^*(\hat{\bm{\Omega}}_k)\rangle_p$ to facilitate a simpler subsequent evaluation. This quantity can be decomposed as
\m
\langle F_A(\hat{\bm{\Omega}}_j)F_B^*(\hat{\bm{\Omega}}_k)\rangle_p=\langle F_A^{+}(\hat{\bm{\Omega}}_j)F_B^{+}(\hat{\bm{\Omega}}_k)
+ F_A^{\times}(\hat{\bm{\Omega}}_j)F_B^{\times}(\hat{\bm{\Omega}}_k)\rangle_p+i \langle F_A^{+}(\hat{\bm{\Omega}}_j)F_B^{\times}(\hat{\bm{\Omega}}_k)
- F_A^{\times}(\hat{\bm{\Omega}}_j)F_B^{+}(\hat{\bm{\Omega}}_k)\rangle_p
\label{FAFB1}.
\n
First, since $\langle F_A(\hat{\bm{\Omega}}_j)F_B^*(\hat{\bm{\Omega}}_k)\rangle_p$ is averaged over a large number of pulsar pairs with identical angular separations, any dependence on the individual pulsar positions is effectively averaged out.
Second, when a pair of GW sources is observed via many pulsar pairs uniformly distributed across the sky, their absolute positions become irrelevant.
Thus, the quantity $\langle F_A(\hat{\bm{\Omega}}_j)F_B^*(\hat{\bm{\Omega}}_k)\rangle_p$ is expected to depend only on the angular separation between the two pulsars, $\gm_{AB} = \cos^{-1}(\hat{\bm{p}}_A \cdot \hat{\bm{p}}_B)$, and the angular separation between the two GW sources, $\beta_{jk} = \cos^{-1} (\hat{\bm{\Omega}}_j \cdot \hat{\bm{\Omega}}_k)$. Accordingly, Eq.~\eqref{FAFB1} can be rewritten as
\m
\langle F_A(\hat{\bm{\Omega}}_j)F_B^*(\hat{\bm{\Omega}}_k)\rangle_p = \mu_{++}(\gm_{AB},\bt_{jk}) + \mu_{\times \times}(\gm_{AB},\bt_{jk}) + i \left[ \mu_{+\times}(\gm_{AB},\bt_{jk}) - \mu_{\times +}(\gm_{AB},\bt_{jk}) \right],
\n
where the shorthand notation is defined as follows:\,
$\mu_{++}(\gm_{AB},\,\bt_{jk}) \,\equiv\, \langle F_A^{+}(\hat{\bm{\Omega}}_j)F_B^{+}(\hat{\bm{\Omega}}_k)\rangle_p$,\,\,\,\,
$\mu_{\times\times} (\gm_{AB},\,\bt_{jk})  \,\equiv\, $ \\$ \langle F_A^{\times}(\hat{\bm{\Omega}}_j)F_B^{\times}(\hat{\bm{\Omega}}_k)\rangle_p$,\,
 $\mu_{+\times} (\gm_{AB},\bt_{jk}) \equiv \langle F_A^{+}(\hat{\bm{\Omega}}_j)F_B^{\times}(\hat{\bm{\Omega}}_k)\rangle_p$,\,
and
$\mu_{\times+}(\gm_{AB},\bt_{jk}) \equiv \langle F_A^{\times}(\hat{\bm{\Omega}}_j)F_B^{+}(\hat{\bm{\Omega}}_k)\rangle_p$.

Furthermore, consider the simultaneous transformation of the pulsar directions from $\hat{\bm{p}}_A \rightarrow -\hat{\bm{p}}_A$ and $\hat{\bm{p}}_B \rightarrow -\hat{\bm{p}}_B$, along with the source directions from $\hat{\bm{\Omega}}_j \rightarrow -\hat{\bm{\Omega}}_j$ and $\hat{\bm{\Omega}}_k \rightarrow -\hat{\bm{\Omega}}_k$. Under this transformation, the angular separations $\gm_{AB}$ and $\bt_{jk}$ remain unchanged, implying that the correlation functions $\mu_{+\times}$ and $\mu_{\times+}$ should also remain invariant.
However, it is known that under such a parity transformation, the plus (``+") polarization mode remains unchanged, while the cross (``$\times$") mode flips sign. Consequently, both $\langle F_A^{+}(\hat{\bm{\Omega}}_j)F_B^{\times}(\hat{\bm{\Omega}}_k)\rangle_p$ and $\langle F_A^{\times}(\hat{\bm{\Omega}}_j)F_B^{+}(\hat{\bm{\Omega}}_k)\rangle_p$ must change sign. The only function that remains unchanged while simultaneously changing its sign is the zero function. Therefore, we conclude that both $\mu_{+\times}$ and $\mu_{\times+}$ must vanish. Accordingly, the ``pulsar-averaged" correlation $\langle \rho_{AB} \rangle_p$ in \Eq{rhoave} simplifies to
\m
\langle\rho_{AB}\rangle_p \!\!\!&=&\!\!\! \frac{1}{2} \sum_{j,k} A_j A_k \cos\(\phi_j - \phi_k\) \langle F_A^{+}(\hat{\Omega}_j)F_B^{+}(\hat{\Omega}_k)+
F_A^{\times}(\hat{\Omega}_j) F_B^{\times}(\hat{\Omega}_k)\rangle_p, \notag \\ 
\!\!\!&=&\!\!\! \frac{1}{2} \sum_{j} A_j^2 \langle F_A^{+}(\hat{\Omega}_j)F_B^{+}(\hat{\Omega}_j)\!+\!
F_A^{\times}(\hat{\Omega}_j) F_B^{\times}(\hat{\Omega}_j)\rangle_p +\frac{1}{2} \sum_{j\neq k} A_j A_k \cos\(\phi_j \!-\! \phi_k\) \langle F_A^{+}(\hat{\Omega}_j)F_B^{+}(\hat{\Omega}_k)\!+\!
F_A^{\times}(\hat{\Omega}_j) F_B^{\times}(\hat{\Omega}_k)\rangle_p\notag\\
\!\!\!&=&\!\!\! \frac{1}{2} \sum_{j} A_j^2 \mu(\gm_{AB})+ \frac{1}{2} \sum_{j\neq k} A_j A_k \cos\(\phi_j \!-\! \phi_k\) \mu(\gm_{AB},\bt_{jk})
\label{rho_AB}
\n
where we have defined
\m
\mu(\gm_{AB}, \beta_{jk}) & \equiv & \langle F_A^{+}(\hat{\Omega}_j)F_B^{+}(\hat{\Omega}_k)+
F_A^{\times}(\hat{\Omega}_j) F_B^{\times}(\hat{\Omega}_k)\rangle_p, \notag\\
\mu(\gm_{AB})&\equiv& \mu(\gm_{AB},0)= \langle F_A^{+}(\hat{\Omega}_j)F_B^{+}(\hat{\Omega}_j)\!+\!
F_A^{\times}(\hat{\Omega}_j) F_B^{\times}(\hat{\Omega}_j)\rangle_p .
\n
Here, the diagonal term $\mu(\gm_{AB})$ can be interpreted as the result of a GW interfering with itself. When the non-diagonal term arising from mutual interference between different sources is neglected, this term is expected to reproduce the standard angular correlation predicted by massive gravity \cite{Liang:2021bct}, up to a normalization factor. This expectation is supported by the interpretation in \cite{Cornish:2013aba}, which demonstrated that a single black hole source produces the same angular correlation pattern as an isotropic stochastic background.
From \Eq{rho_AB}, we extract the overlap reduction function (ORF) that encapsulates the geometric dependence as
\e
\Gamma(\gm_{AB})=\chi\[\mu(\gm_{AB})+ \sum_{j \neq k} \mathcal{A}_{j}\mathcal{A}_{k} \cos\(\phi_j - \phi_k\) \mu(\gm_{AB}, \beta_{jk})\],
\label{gammanorm}
\q 
where $\mathcal{A}_{j}\mathcal{A}_{k}=A_j A_k/\sum_{n}A_n^2$ and $\chi$ is a normalization constant.
Since the ORF characterizes the redshift response correlation between two pulsars, it is appropriate for the correlation coefficient to equal unity when the two pulsars are identical (i.e., $A = B$) \cite{Romano:2023zhb}, consistent with the standard HD prescription commonly used in signal searches that neglect mutual interference \cite{NANOGrav:2023gor}. Accordingly, $\chi$ is determined by requiring $\Gamma(\gm_{AA}) = 1$ for auto-correlated pulsars.  From \Eq{rhoave}, we find that the $\Gm(\gm_{AA})$ should be written as
\m
\Gamma(\gm_{AA})=\chi\[2\mu(0)+ \sum_{j \neq k} \mathcal{A}_{j}\mathcal{A}_{k} \cos\(\phi_j - \phi_k\) \mu(0, \beta_{jk})\],
\label{auto_gamma}
\n
as the exponential term $\mathrm{e}^{-2\pi i [f_j L_A (1 + \eta \hat{\bm{\Omega}}_j \cdot \hat{\bm{p}}_A)-f_k L_A (1 + \eta \hat{\bm{\Omega}}_k \cdot \hat{\bm{p}}_A)]}$ introduces an additional contribution $\dt_{jk}$. Therefore, the normalized cross correlation function should take the form
\m
\Gamma(\gm_{AB})=\frac{1}{2\mu(0)+ \sum_{j \neq k} \mathcal{A}_{j}\mathcal{A}_{k} \cos\(\phi_j - \phi_k\) \mu(0, \beta_{jk})}\[\mu(\gm_{AB})+ \sum_{j \neq k} \mathcal{A}_{j}\mathcal{A}_{k} \cos\(\phi_j - \phi_k\) \mu(\gm_{AB}, \beta_{jk})\],
\label{norm_Gm}
\n
which accounts for both auto-interference from individual GW sources and mutual interference between different sources.

\section{Spatial Correlation Features in Massive Gravity}
We now proceed to compute the two-point function $\mu(\gm_{AB}, \beta_{jk})$. For brevity and without loss of clarity, we occasionally omit the subscripts of $\gm_{AB}$ and $\bt_{jk}$, referring to them simply as $\gm$ and $\bt$, respectively.
Following the technique in \cite{Allen:2022dzg}, we begin by choosing separate coordinate systems for the two GW sources as follows:
\begin{align}
\text{GW $j$:} \quad
\left\{
\begin{aligned}
\hat{\Omega}_j &= (\sin \beta, \, 0, \,\cos \beta) \\
\hat{m}_j &= (0,\, -1,\, 0) \\
\hat{n}_j &= (\cos \beta, \,0,\, -\sin \beta)
\end{aligned}
\right. 
\qquad \qquad \qquad
\text{GW $k$:} \quad
\left\{
\begin{aligned}
\hat{\Omega}_k &= (0,\, 0,\, 1) \\
\hat{m}_k &= (0,\, -1,\, 0) \\
\hat{n}_k &= (1,\, 0,\, 0)
\end{aligned}
\right. \qquad,
\label{GWxyz}
\end{align}
from which the polarization tensors of each GW source can be obtained according to
\m
\label{e_ab}
\hat{\varepsilon}^{+}_{ab}=\hat{m}_a \hat{m}_b- \hat{n}_a \hat{n}_b,\\ \notag
\hat{\varepsilon}^{\times}_{ab}=\hat{m}_a \hat{n}_b+ \hat{n}_a \hat{m}_b.
\n
We then locate the two pulsars at 
\m
\label{Pxyz}
\hat{p}_A&=&(\sin \theta \cos \lbd, \,\, \sin \theta \sin \lbd, \,\, \cos \theta),\\ \notag
\hat{p}_B&=&\(\cos \lambda (\cos \gamma \sin \theta-\sin \gamma  \cos \theta  \cos \phi )+\sin \gamma  \sin \lambda  \sin \phi , \right. \\ \notag 
 &\,&\,\, \left. \sin \lambda  (\cos \gamma  \sin \theta -\sin \gamma  \cos \theta \cos \phi )-\sin \gamma  \cos \lambda  \sin \phi , \,\, \sin \gamma  \sin \theta  \cos \phi +\cos \gamma  \cos \theta \),
\n
where $\theta$, $\lbd$ and $\phi$ are spherical coordinate parameters that allow for arbitrary pulsar positions, provided that their angular separation remains $\gm$.
Using \Eq{F+x}, \Eq{GWxyz}, \Eq{e_ab} and \Eq{Pxyz}, it is easy to express $F_A^{+}(\hat{\Omega}_j)F_B^{+}(\hat{\Omega}_k)+
F_A^{\times}(\hat{\Omega}_j) F_B^{\times}(\hat{\Omega}_k)$.

\begin{figure}[htbp]
	\centering	\includegraphics[width=1.\textwidth]{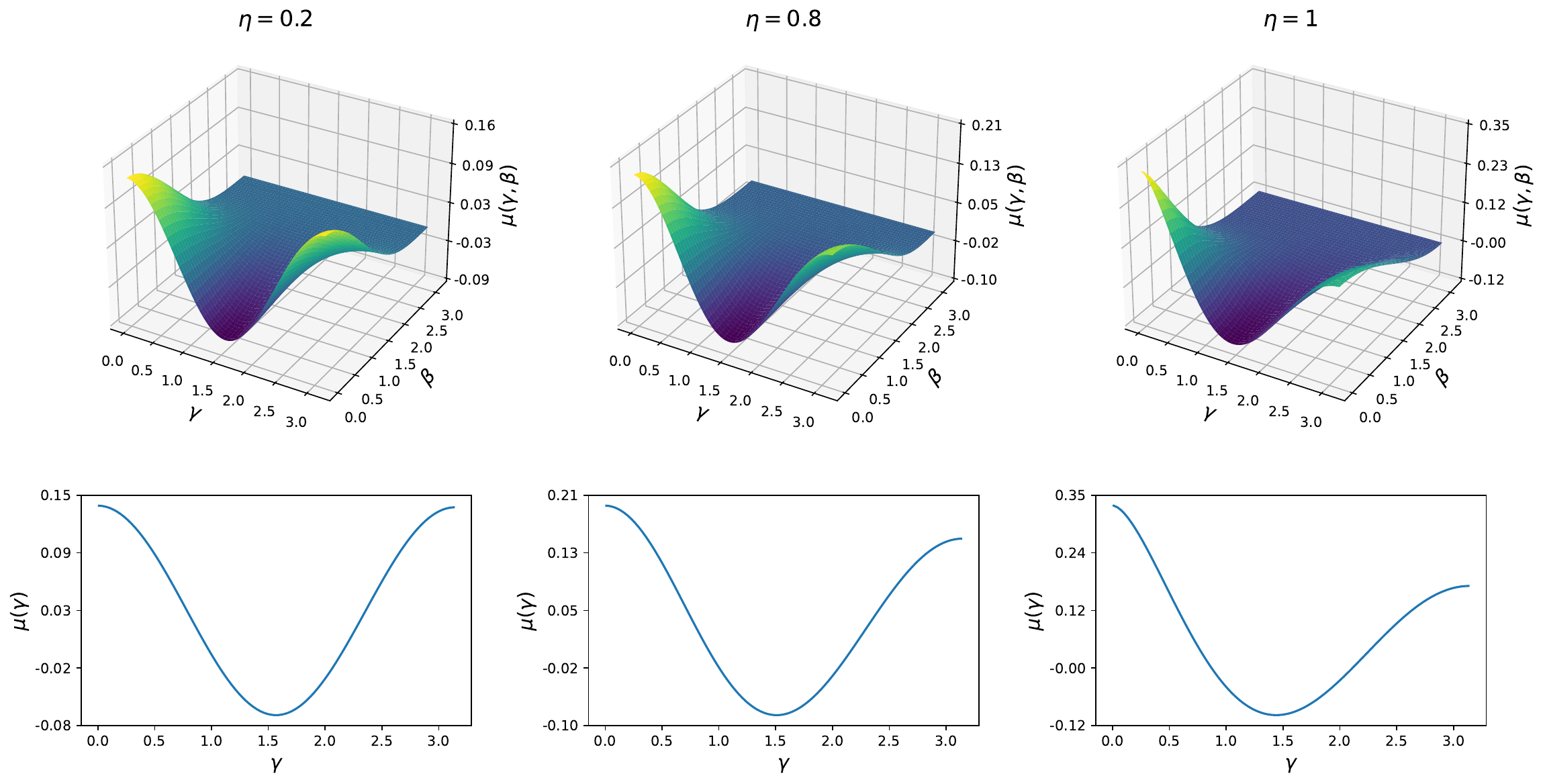}\caption{\textbf{Upper panel:} The two-point function $\mu(\gm,\bt)$ calculated using \Eq{mu_int}, representing the mean correlation between two pulsars separated by an angle $\gm$, arising from two interfering sources with angular separation $\bt$. Three representative cases are shown for $\eta = 0.2$, $0.8$, and $1$.
 \textbf{Lower panel}:The corresponding correlation $\mu(\gm)$ obtained from \Eq{mu_gm} by setting $\beta = 0$ in $\mu(\gm,\bt)$. This recovers the standard angular correlation when only auto-interference is considered. The same three cases, $\eta = 0.2$, $0.8$, and $1$, are shown.
}
\label{mu_bt_gm}
\end{figure}

The next step is to perform the pulsar average over $F_A^{+}(\hat{\Omega}_j)F_B^{+}(\hat{\Omega}_k)+
F_A^{\times}(\hat{\Omega}_j) F_B^{\times}(\hat{\Omega}_k)$, which can be carried out by sequentially integrating over the angular variables: first $\lbd$ over the range $[0, 2\pi]$, then $\phi$ over $[0, 2\pi]$, and finally $\theta$ over $[0, \pi]$, i.e.,
\m
\mu(\gm,\bt)=\frac{1}{8\pi^2}\int_0^{\pi} \sin \theta \,{\rm{d}\theta} \int_0^{2\pi} {\rm{d}\phi} \int_0^{2\pi} {\rm{d}\lbd} \,\[F_A^{+}(\hat{\Omega}_j)F_B^{+}(\hat{\Omega}_k)+
F_A^{\times}(\hat{\Omega}_j) F_B^{\times}(\hat{\Omega}_k)\].  
\n
After completing the first two integrations, we change the integration variable from $\theta$ to $x$ via the substitution $x\equiv \cos \theta$ and obtain
\m
\label{mu_int}
\mu(\gm,\bt)\!=\!\int_{-1}^{+1} {\rm{d}} x \left\{\frac{\csc ^4\beta }{128 \eta^6 \left(x^2-1\right)^2}\right. \!\!\!\! \!\!\!\!\!&&\left[\left(-\sqrt{\eta^2 \cos ^2\beta+\eta^2 \left(x^2-1\right)+2 \eta x \cos \beta +1}+\eta x \cos \beta +1\right)^2\right.\\ \notag
&&\left. \left(\frac{\left(\eta^2+1\right) \cos 2 \beta +4 \eta^2 x^2-\eta^2+8 \eta x \cos \beta +3}{\sqrt{\eta^2 \cos ^2\beta +\eta^2 \left(x^2-1\right)+2 \eta x \cos \beta +1}}+4 (\cos \beta +\eta x)\right) \right. \\ \notag 
&& \left. \left. \left(\frac{8 \left(\eta^2 \cos 2 \gamma +\eta^2 x^2+4 \eta x \cos \gamma +x^2+1\right)}{\sqrt{\eta^2 \cos ^2\gamma +\eta^2 \left(x^2-1\right)+2 \eta x \cos \gamma +1}}+8 \eta x \left(x^2-3\right) \cos \gamma -8 \left(x^2+1\right)\right)\right]\right\}.
\n
The integral expression for $\mu(\gamma,\beta)$ can be conveniently evaluated numerically for a given $\eta$. Several representative examples with $\eta=0.2, 0.8$ and $1$ are shown in the upper panel of \Fig{mu_bt_gm}. In particular, when $\beta = 0$, the integral simplifies to 
\begin{align}
\label{mu_gm}
\mu(\gm)=-\frac{1}{24 \eta^5} \Bigg\{
  & 2 \eta \left(-3 + (-6 + 5 \eta^2) \cos\gamma \right)
  + 6 \left(1 + \eta^2 + (1 - 3 \eta^2) \cos\gamma \right)
  \log\left(\frac{1 + \eta}{1 - \eta}\right) \\
  & + \frac{3 \left(1 + 2 \eta^2 - 4 \eta^2 \cos\gamma + \eta^4 \cos(2\gamma)\right)}{
  \sqrt{(1 + \cos\gamma)(2 - \eta^2 - \eta^2 \cos\gamma)}} \notag \\
  & \times \log\Bigg[
    \frac{\Big(
      1 + 2 \eta^2 (1 - 2 \cos\gamma)
      - 2 \eta (1 - \eta^2 \cos\gamma)
      \sqrt{(1 - \cos\gamma)(2 - \eta^2 - \eta^2 \cos\gamma)}  + \eta^4 \cos(2\gamma)
    \Big)}{(1 - \eta^2)^2} 
  \Bigg]
\Bigg\},\notag
\end{align}
which exactly reproduces the conventional angular correlation form derived in the absence of GW source interference \cite{Liang:2021bct}. Examples corresponding to this special case are presented in the lower panel of \Fig{mu_bt_gm}.
It is evident that smaller values of $\eta$, or equivalently, larger graviton masses at fixed frequency (according to \Eq{eta}), lead to an auto-interference correlation curve that becomes increasingly symmetric. 

While the auto-interference term leads to a well-defined and predictable structure, the contribution from mutual interference between GW sources is inherently unpredictable, as it depends on source-specific information such as individual phases and sky locations—quantities that are generally inaccessible. As a result, the full correlation curve given in \Eq{norm_Gm} also becomes unpredictable. To explore its statistical properties, we therefore turn to numerical simulations.

\begin{figure}[htpb]
	\centering
 \includegraphics[width=1.\textwidth]{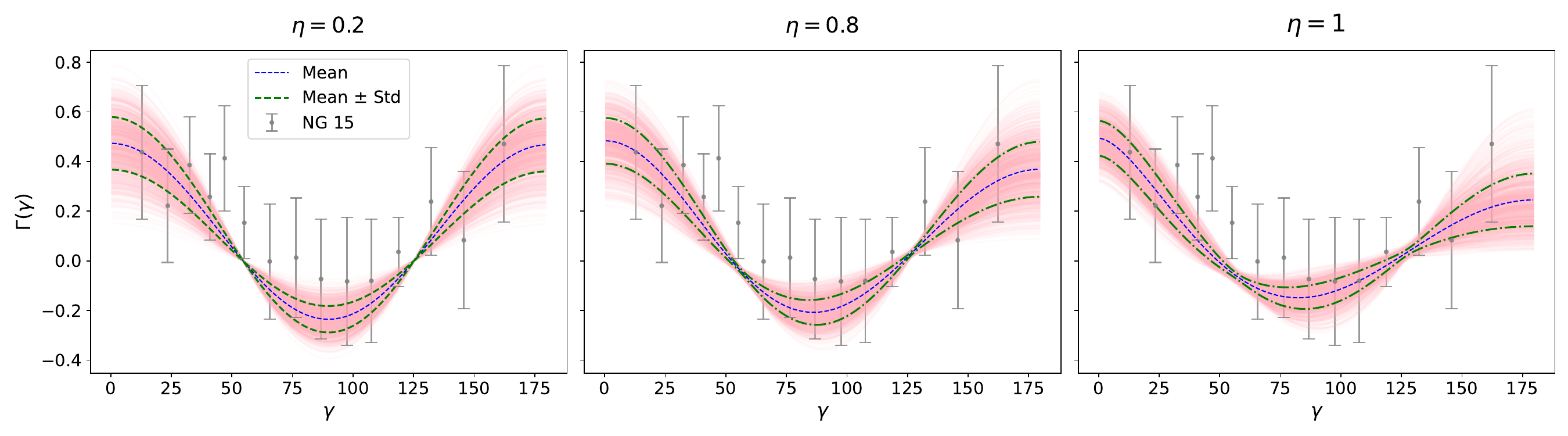}\caption{
 Normalized correlation curves $\Gm(\gm)$ from 1,000 realizations for $\eta = 0.2$, $0.8$, and $1$. Pink curves represent individual realizations, the blue dotted line indicates the mean correlation, and the green dashed lines denote the one-standard-deviation range around the mean. For comparison, the grey error bars depict the angular-separation-binned inter-pulsar correlations measured in the NANOGrav 15-year data set (NG 15) using the frequentist optimal statistic.
 \label{orf_shape}}
\end{figure}

We generate 1,000 realizations of the correlation for the representative cases with $\eta = 0.2$, $0.8$, and $1$. In each realization, 1,000 GW sources are uniformly distributed over the sky with random directions $\Omega_j$ and random phases $\phi_j$. The amplitudes $A_j$ are drawn from a uniform distribution over the interval [0.01, 1]\footnote{We have also tested alternative sampling methods for the amplitudes—for example, setting all $A_j = 1$, or assuming a uniform spatial density of sources corresponding to $A_j = j^{1/3}$~\cite{Allen:2022dzg}. These variations yield similar results.}. The different interference configurations arising from various combinations of GW source positions and phases in each realization result in different correlations according to \Eq{norm_Gm}, as shown in \Fig{orf_shape}. The pink area consists of individual pink curves, each representing the correlation from a single realization;
the blue dotted line shows the mean correlation across all realizations, and the green dashed lines indicate the region within one standard deviation of the mean. This standard deviation, calculated from an ensemble of realizations and stemming from fluctuations induced by source interference, is commonly referred to as the ``cosmic variance”~\cite{Allen:2022dzg}. It is important to emphasize, however, that we can observe only a single realization of our unique Universe.

\begin{figure}[H]
	\centering
 \includegraphics[width=0.75\textwidth]{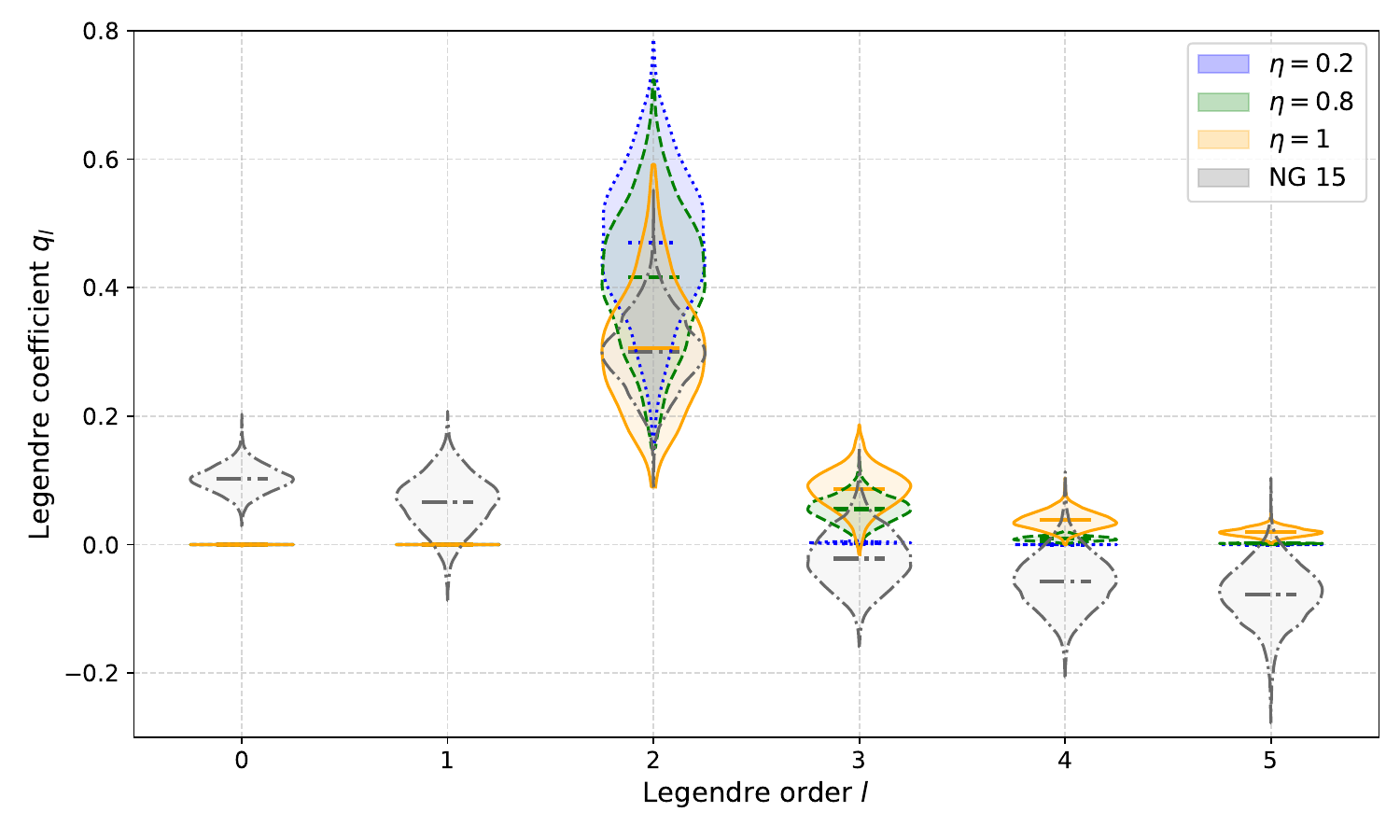}\caption{Violin plots of the coefficients $q_l$ from the Legendre polynomial decomposition \Eq{mu_ql} of the full normalized correlation $\Gm(\gm)$, shown up to the fifth order and based on 1,000 simulations. The purple violins with dotted outlines, green violins with dashed outlines, and yellow violins with solid outlines correspond to the cases of $\eta = 0.2$, $0.8$, and $1$, respectively. Horizontal lines indicate the means of the distributions. For comparison, the grey violins with dash-dotted outlines depict the distribution of the Legendre coefficients measured in the NANOGrav 15-year data set (NG 15) using the multiple-component optimal statistic. \label{orf_err_bar}}
\end{figure}

To investigate the detailed structure of the full correlation, we expand each curve in terms of Legendre polynomials as
\m
\Gamma(\gm) = \sum_{l} q_l P_l(\cos \gm),
\label{mu_ql}
\n
where $P_l(\cos \gm)$ denotes the Legendre polynomial of order $l$ evaluated at angle $\gm$, and $q_l$ are the corresponding Legendre coefficients, which vary from realization to realization. Violin plots illustrating the distributions of $q_l$ up to the fifth order are presented in \Fig{orf_err_bar}. We find that, overall, the monopole ($l=0$) and dipole ($l=1$) components contribute negligibly, while significant contributions emerge at $l \ge 2$ but diminish with increasing $l$. Moreover, the dominance of the quadrupole ($l=2$) component becomes increasingly pronounced as $\eta$ decreases, consistent with the more symmetric trend observed in \Fig{orf_shape}. These features reflect both the intrinsic quadrupolar nature and the effects of the modified dispersion relation of GW radiation. Similar plots were presented in the top panel of Fig. 2 in~\cite{Bernardo:2023pwt}, which were obtained using Gaussian ensemble calculations~\cite{Bernardo:2022xzl}.

\FloatBarrier

\section{Summary and Discussion}

In the nanohertz band, numerous GW sources emitting at similar frequencies interfere with one another. As a result, the spatial correlation pattern induced by a realistic SGWB in PTAs is shaped by such interference, deviating from the conventional correlation derived under the assumption of an isotropic Gaussian ensemble of sources.

In this work, we investigate how GW interference affects the spatial correlation function in the context of massive gravity. Through simulations, we demonstrate that the resulting angular correlations—determined by specific interference configurations—are well characterized by a Legendre polynomial expansion. These correlations predominantly receive contributions from modes with $l \ge 2$, with the coefficients generally decreasing as $l$ increases. Moreover, as the graviton mass increases (for a fixed frequency), the quadrupole component ($l=2$) becomes increasingly dominant. These trends reflect both the quadrupolar nature intrinsic to GW radiation and the modified dispersion relation arising in massive gravity.

While it has been anticipated that spatial correlation measurements in PTAs could help distinguish massive gravity from GR, GW interference poses a significant complication. Since we can observe only a single Universe—with a fixed (albeit unknown) configuration of GW source positions and phases—we cannot rely on statistical averaging over multiple realizations to discriminate between theories. On the other hand, many possible interference configurations can produce angular correlation patterns that are consistent with either theory (also broadly compatible with the NANOGrav 15-year measurements; see \Fig{orf_shape}), making it difficult to draw a clear distinction from a single realization.

We now consider this issue in greater detail. In PTA data analysis, an SGWB is characterized by the cross-power spectral density $S_{AB}(f) = P(f)\Gamma(\gamma_{AB})$ at discrete Fourier frequencies $f_i = i/T$ ($i=1,2,\cdots$), where $P(f)$ is the power spectral density of the GWB. In principle, both $P(f)$ and $\Gamma(\gamma_{AB})$ can provide insights into physics consistent or beyond GR.

From \Eq{eta}, we note that a nonzero graviton mass implies a minimum allowed frequency for GWs, given by
\m
f_{\text{min}}=\frac{m_g}{2\pi}.
\n
For a PTA operating over a timespan $T$, the absence of power in $P(f)$ below a certain frequency may indicate the presence of a graviton mass. Conversely, if $P(f)$ remains nonzero at the lowest accessible frequency $1/T$, this sets a fundamental upper bound on the graviton mass,
\m
m_g \le m_b = \frac{2\pi}{T}.
\n

One might consider tightening this constraint using angular correlation information—for example, imposing $m_g < 1/2 \,m_b$. This case, $m_g =1/2\, m_b$, corresponds to values of $\eta = 0.866$, $0.968$, $0.986$, $0.992$, and $0.995$ in the first five frequency bins, respectively. However, since $\eta$ increases with frequency and asymptotically approaches unity in the massless limit, averaging angular correlations across frequency bins—as is valid in GR \cite{Allen:2024uqs}—is not justified. Instead, constraints must rely on individual frequency bins, particularly the lowest one, which exhibits the strongest deviation from $\eta = 1$.

Nevertheless, even in the cases of $\eta = 0.8$ and $\eta = 1$, our simulations, as shown in \Fig{orf_err_bar}, clearly indicate that their Legendre coefficients largely overlap with each other for the dominant $l=2$ and $l=3$ modes. 
This suggests that the ability to distinguish between these two scenarios is very limited, even under ideal, noise-free conditions; although, in principle, resolving the higher-order modes at a precision of $\mathcal{O}(10^{-3})$ could provide a way to discriminate between them~\citep{Bernardo:2023pwt}. In realistic observational settings, noise further degrades the precision with which these coefficients can be measured, making it even more challenging to resolve such subtle differences. For instance, the NANOGrav 15-year results (Figure 7 of \cite{NANOGrav:2023gor}), reproduced in our \Fig{orf_err_bar} as grey violin plots with dash-dotted outlines, exhibit large uncertainties across all modes from $l=0$ to $5$. The apparent monopole and dipole components remain of uncertain origin and are not strongly supported by complementary statistical analyses \cite{NANOGrav:2023gor}. Moreover, the $l\ge 2$ coefficients—despite negative median values for high-order modes—are consistent with all three representative cases, $\eta=0.2, 0.8,$ and $1$, within the range of uncertainties, indicating that each could serve as a viable explanation.

Therefore, once realistic GW interference from different sources is accounted for, the angular correlation function is unlikely to yield a substantial improvement in the graviton mass constraint beyond the spectral bound set by the lowest observable frequency.

It is important to note, however, that such interference effects are primarily relevant for GW backgrounds of astrophysical origin. In contrast, for a cosmological GW background—such as those produced by first-order phase transitions or cosmic strings—the ergodic hypothesis applies~\cite{Caprini:2018mtu}. The reason is that the present-day GWB results from the superposition of many independent signals emitted by uncorrelated regions in the early Universe. As a result, we effectively access multiple realizations and can observe a predictable, ensemble-averaged correlation pattern, which theoretically retains distinctive features capable of discerning a massive graviton.

\section*{Data Availability Statement}
The simulation data used in this study are available from ~\cite{massive_data}. 
\begin{acknowledgments}
 
QGH is supported by the National Key Research and Development Program of China Grant No.2020YFC2201502, the National Natural
Science Foundation of China (Grant No.~12475065, 12447101) and the China Manned Space Program with grant no. CMS-CSST-2025-A01. YMW is supported by the National Natural
Science Foundation of China (Grant No.~12505086).

\end{acknowledgments}


\bibliography{refs}	

\begin{thebibliography}{56}%
\makeatletter
\providecommand \@ifxundefined [1]{%
 \@ifx{#1\undefined}
}%
\providecommand \@ifnum [1]{%
 \ifnum #1\expandafter \@firstoftwo
 \else \expandafter \@secondoftwo
 \fi
}%
\providecommand \@ifx [1]{%
 \ifx #1\expandafter \@firstoftwo
 \else \expandafter \@secondoftwo
 \fi
}%
\providecommand \natexlab [1]{#1}%
\providecommand \enquote  [1]{``#1''}%
\providecommand \bibnamefont  [1]{#1}%
\providecommand \bibfnamefont [1]{#1}%
\providecommand \citenamefont [1]{#1}%
\providecommand \href@noop [0]{\@secondoftwo}%
\providecommand \href [0]{\begingroup \@sanitize@url \@href}%
\providecommand \@href[1]{\@@startlink{#1}\@@href}%
\providecommand \@@href[1]{\endgroup#1\@@endlink}%
\providecommand \@sanitize@url [0]{\catcode `\\12\catcode `\$12\catcode
  `\&12\catcode `\#12\catcode `\^12\catcode `\_12\catcode `\%12\relax}%
\providecommand \@@startlink[1]{}%
\providecommand \@@endlink[0]{}%
\providecommand \url  [0]{\begingroup\@sanitize@url \@url }%
\providecommand \@url [1]{\endgroup\@href {#1}{\urlprefix }}%
\providecommand \urlprefix  [0]{URL }%
\providecommand \Eprint [0]{\href }%
\providecommand \doibase [0]{http://dx.doi.org/}%
\providecommand \selectlanguage [0]{\@gobble}%
\providecommand \bibinfo  [0]{\@secondoftwo}%
\providecommand \bibfield  [0]{\@secondoftwo}%
\providecommand \translation [1]{[#1]}%
\providecommand \BibitemOpen [0]{}%
\providecommand \bibitemStop [0]{}%
\providecommand \bibitemNoStop [0]{.\EOS\space}%
\providecommand \EOS [0]{\spacefactor3000\relax}%
\providecommand \BibitemShut  [1]{\csname bibitem#1\endcsname}%
\let\auto@bib@innerbib\@empty
\bibitem [{\citenamefont {Fierz}\ and\ \citenamefont
  {Pauli}(1939)}]{Fierz:1939ix}%
  \BibitemOpen
  \bibfield  {author} {\bibinfo {author} {\bibfnamefont {M.}~\bibnamefont
  {Fierz}}\ and\ \bibinfo {author} {\bibfnamefont {W.}~\bibnamefont {Pauli}},\
  }\bibfield  {title} {\enquote {\bibinfo {title} {{On relativistic wave
  equations for particles of arbitrary spin in an electromagnetic field}},}\
  }\href {\doibase 10.1098/rspa.1939.0140} {\bibfield  {journal} {\bibinfo
  {journal} {Proc. Roy. Soc. Lond. A}\ }\textbf {\bibinfo {volume} {173}},\
  \bibinfo {pages} {211--232} (\bibinfo {year} {1939})}\BibitemShut {NoStop}%
\bibitem [{\citenamefont {van Dam}\ and\ \citenamefont
  {Veltman}(1970)}]{vanDam:1970vg}%
  \BibitemOpen
  \bibfield  {author} {\bibinfo {author} {\bibfnamefont {H.}~\bibnamefont {van
  Dam}}\ and\ \bibinfo {author} {\bibfnamefont {M.~J.~G.}\ \bibnamefont
  {Veltman}},\ }\bibfield  {title} {\enquote {\bibinfo {title} {{Massive and
  massless Yang-Mills and gravitational fields}},}\ }\href {\doibase
  10.1016/0550-3213(70)90416-5} {\bibfield  {journal} {\bibinfo  {journal}
  {Nucl. Phys. B}\ }\textbf {\bibinfo {volume} {22}},\ \bibinfo {pages}
  {397--411} (\bibinfo {year} {1970})}\BibitemShut {NoStop}%
\bibitem [{\citenamefont {Zakharov}(1970)}]{Zakharov:1970cc}%
  \BibitemOpen
  \bibfield  {author} {\bibinfo {author} {\bibfnamefont {V.~I.}\ \bibnamefont
  {Zakharov}},\ }\bibfield  {title} {\enquote {\bibinfo {title} {{Linearized
  gravitation theory and the graviton mass}},}\ }\href@noop {} {\bibfield
  {journal} {\bibinfo  {journal} {JETP Lett.}\ }\textbf {\bibinfo {volume}
  {12}},\ \bibinfo {pages} {312} (\bibinfo {year} {1970})}\BibitemShut
  {NoStop}%
\bibitem [{\citenamefont {Vainshtein}(1972)}]{Vainshtein:1972sx}%
  \BibitemOpen
  \bibfield  {author} {\bibinfo {author} {\bibfnamefont {A.~I.}\ \bibnamefont
  {Vainshtein}},\ }\bibfield  {title} {\enquote {\bibinfo {title} {{To the
  problem of nonvanishing gravitation mass}},}\ }\href {\doibase
  10.1016/0370-2693(72)90147-5} {\bibfield  {journal} {\bibinfo  {journal}
  {Phys. Lett. B}\ }\textbf {\bibinfo {volume} {39}},\ \bibinfo {pages}
  {393--394} (\bibinfo {year} {1972})}\BibitemShut {NoStop}%
\bibitem [{\citenamefont {Boulware}\ and\ \citenamefont
  {Deser}(1972)}]{Boulware:1972yco}%
  \BibitemOpen
  \bibfield  {author} {\bibinfo {author} {\bibfnamefont {D.~G.}\ \bibnamefont
  {Boulware}}\ and\ \bibinfo {author} {\bibfnamefont {Stanley}\ \bibnamefont
  {Deser}},\ }\bibfield  {title} {\enquote {\bibinfo {title} {{Can gravitation
  have a finite range?}}}\ }\href {\doibase 10.1103/PhysRevD.6.3368} {\bibfield
   {journal} {\bibinfo  {journal} {Phys. Rev. D}\ }\textbf {\bibinfo {volume}
  {6}},\ \bibinfo {pages} {3368--3382} (\bibinfo {year} {1972})}\BibitemShut
  {NoStop}%
\bibitem [{\citenamefont {Dvali}\ and\ \citenamefont
  {Gabadadze}(2001)}]{Dvali:2000xg}%
  \BibitemOpen
  \bibfield  {author} {\bibinfo {author} {\bibfnamefont {G.~R.}\ \bibnamefont
  {Dvali}}\ and\ \bibinfo {author} {\bibfnamefont {Gregory}\ \bibnamefont
  {Gabadadze}},\ }\bibfield  {title} {\enquote {\bibinfo {title} {{Gravity on a
  brane in infinite volume extra space}},}\ }\href {\doibase
  10.1103/PhysRevD.63.065007} {\bibfield  {journal} {\bibinfo  {journal} {Phys.
  Rev. D}\ }\textbf {\bibinfo {volume} {63}},\ \bibinfo {pages} {065007}
  (\bibinfo {year} {2001})},\ \Eprint {http://arxiv.org/abs/hep-th/0008054}
  {arXiv:hep-th/0008054} \BibitemShut {NoStop}%
\bibitem [{\citenamefont {Dvali}\ \emph
  {et~al.}(2000{\natexlab{a}})\citenamefont {Dvali}, \citenamefont
  {Gabadadze},\ and\ \citenamefont {Porrati}}]{Dvali:2000hr}%
  \BibitemOpen
  \bibfield  {author} {\bibinfo {author} {\bibfnamefont {G.~R.}\ \bibnamefont
  {Dvali}}, \bibinfo {author} {\bibfnamefont {Gregory}\ \bibnamefont
  {Gabadadze}}, \ and\ \bibinfo {author} {\bibfnamefont {Massimo}\ \bibnamefont
  {Porrati}},\ }\bibfield  {title} {\enquote {\bibinfo {title} {{4-D gravity on
  a brane in 5-D Minkowski space}},}\ }\href {\doibase
  10.1016/S0370-2693(00)00669-9} {\bibfield  {journal} {\bibinfo  {journal}
  {Phys. Lett. B}\ }\textbf {\bibinfo {volume} {485}},\ \bibinfo {pages}
  {208--214} (\bibinfo {year} {2000}{\natexlab{a}})},\ \Eprint
  {http://arxiv.org/abs/hep-th/0005016} {arXiv:hep-th/0005016} \BibitemShut
  {NoStop}%
\bibitem [{\citenamefont {Dvali}\ \emph
  {et~al.}(2000{\natexlab{b}})\citenamefont {Dvali}, \citenamefont
  {Gabadadze},\ and\ \citenamefont {Porrati}}]{Dvali:2000rv}%
  \BibitemOpen
  \bibfield  {author} {\bibinfo {author} {\bibfnamefont {G.~R.}\ \bibnamefont
  {Dvali}}, \bibinfo {author} {\bibfnamefont {G.}~\bibnamefont {Gabadadze}}, \
  and\ \bibinfo {author} {\bibfnamefont {M.}~\bibnamefont {Porrati}},\
  }\bibfield  {title} {\enquote {\bibinfo {title} {{Metastable gravitons and
  infinite volume extra dimensions}},}\ }\href {\doibase
  10.1016/S0370-2693(00)00631-6} {\bibfield  {journal} {\bibinfo  {journal}
  {Phys. Lett. B}\ }\textbf {\bibinfo {volume} {484}},\ \bibinfo {pages}
  {112--118} (\bibinfo {year} {2000}{\natexlab{b}})},\ \Eprint
  {http://arxiv.org/abs/hep-th/0002190} {arXiv:hep-th/0002190} \BibitemShut
  {NoStop}%
\bibitem [{\citenamefont {de~Rham}\ \emph {et~al.}(2011)\citenamefont
  {de~Rham}, \citenamefont {Gabadadze},\ and\ \citenamefont
  {Tolley}}]{deRham:2010kj}%
  \BibitemOpen
  \bibfield  {author} {\bibinfo {author} {\bibfnamefont {Claudia}\ \bibnamefont
  {de~Rham}}, \bibinfo {author} {\bibfnamefont {Gregory}\ \bibnamefont
  {Gabadadze}}, \ and\ \bibinfo {author} {\bibfnamefont {Andrew~J.}\
  \bibnamefont {Tolley}},\ }\bibfield  {title} {\enquote {\bibinfo {title}
  {{Resummation of Massive Gravity}},}\ }\href {\doibase
  10.1103/PhysRevLett.106.231101} {\bibfield  {journal} {\bibinfo  {journal}
  {Phys. Rev. Lett.}\ }\textbf {\bibinfo {volume} {106}},\ \bibinfo {pages}
  {231101} (\bibinfo {year} {2011})},\ \Eprint {http://arxiv.org/abs/1011.1232}
  {arXiv:1011.1232 [hep-th]} \BibitemShut {NoStop}%
\bibitem [{\citenamefont {Bernus}\ \emph {et~al.}(2020)\citenamefont {Bernus},
  \citenamefont {Minazzoli}, \citenamefont {Fienga}, \citenamefont {Gastineau},
  \citenamefont {Laskar}, \citenamefont {Deram},\ and\ \citenamefont
  {Di~Ruscio}}]{Bernus:2020szc}%
  \BibitemOpen
  \bibfield  {author} {\bibinfo {author} {\bibfnamefont {L.}~\bibnamefont
  {Bernus}}, \bibinfo {author} {\bibfnamefont {O.}~\bibnamefont {Minazzoli}},
  \bibinfo {author} {\bibfnamefont {A.}~\bibnamefont {Fienga}}, \bibinfo
  {author} {\bibfnamefont {M.}~\bibnamefont {Gastineau}}, \bibinfo {author}
  {\bibfnamefont {J.}~\bibnamefont {Laskar}}, \bibinfo {author} {\bibfnamefont
  {P.}~\bibnamefont {Deram}}, \ and\ \bibinfo {author} {\bibfnamefont
  {A.}~\bibnamefont {Di~Ruscio}},\ }\bibfield  {title} {\enquote {\bibinfo
  {title} {{Constraint on the Yukawa suppression of the Newtonian potential
  from the planetary ephemeris INPOP19a}},}\ }\href {\doibase
  10.1103/PhysRevD.102.021501} {\bibfield  {journal} {\bibinfo  {journal}
  {Phys. Rev. D}\ }\textbf {\bibinfo {volume} {102}},\ \bibinfo {pages}
  {021501} (\bibinfo {year} {2020})},\ \Eprint
  {http://arxiv.org/abs/2006.12304} {arXiv:2006.12304 [gr-qc]} \BibitemShut
  {NoStop}%
\bibitem [{\citenamefont {Brito}\ \emph {et~al.}(2013)\citenamefont {Brito},
  \citenamefont {Cardoso},\ and\ \citenamefont {Pani}}]{Brito:2013wya}%
  \BibitemOpen
  \bibfield  {author} {\bibinfo {author} {\bibfnamefont {Richard}\ \bibnamefont
  {Brito}}, \bibinfo {author} {\bibfnamefont {Vitor}\ \bibnamefont {Cardoso}},
  \ and\ \bibinfo {author} {\bibfnamefont {Paolo}\ \bibnamefont {Pani}},\
  }\bibfield  {title} {\enquote {\bibinfo {title} {{Massive spin-2 fields on
  black hole spacetimes: Instability of the Schwarzschild and Kerr solutions
  and bounds on the graviton mass}},}\ }\href {\doibase
  10.1103/PhysRevD.88.023514} {\bibfield  {journal} {\bibinfo  {journal} {Phys.
  Rev. D}\ }\textbf {\bibinfo {volume} {88}},\ \bibinfo {pages} {023514}
  (\bibinfo {year} {2013})},\ \Eprint {http://arxiv.org/abs/1304.6725}
  {arXiv:1304.6725 [gr-qc]} \BibitemShut {NoStop}%
\bibitem [{\citenamefont {Choudhury}\ \emph {et~al.}(2004)\citenamefont
  {Choudhury}, \citenamefont {Joshi}, \citenamefont {Mahajan},\ and\
  \citenamefont {McKellar}}]{Choudhury:2002pu}%
  \BibitemOpen
  \bibfield  {author} {\bibinfo {author} {\bibfnamefont {S.~R.}\ \bibnamefont
  {Choudhury}}, \bibinfo {author} {\bibfnamefont {Girish~C.}\ \bibnamefont
  {Joshi}}, \bibinfo {author} {\bibfnamefont {S.}~\bibnamefont {Mahajan}}, \
  and\ \bibinfo {author} {\bibfnamefont {Bruce H.~J.}\ \bibnamefont
  {McKellar}},\ }\bibfield  {title} {\enquote {\bibinfo {title} {{Probing large
  distance higher dimensional gravity from lensing data}},}\ }\href {\doibase
  10.1016/j.astropartphys.2004.04.001} {\bibfield  {journal} {\bibinfo
  {journal} {Astropart. Phys.}\ }\textbf {\bibinfo {volume} {21}},\ \bibinfo
  {pages} {559--563} (\bibinfo {year} {2004})},\ \Eprint
  {http://arxiv.org/abs/hep-ph/0204161} {arXiv:hep-ph/0204161} \BibitemShut
  {NoStop}%
\bibitem [{\citenamefont {Abbott}\ \emph {et~al.}(2016)\citenamefont {Abbott}
  \emph {et~al.}}]{LIGOScientific:2016lio}%
  \BibitemOpen
  \bibfield  {author} {\bibinfo {author} {\bibfnamefont {B.~P.}\ \bibnamefont
  {Abbott}} \emph {et~al.} (\bibinfo {collaboration} {LIGO Scientific,
  Virgo}),\ }\bibfield  {title} {\enquote {\bibinfo {title} {{Tests of general
  relativity with GW150914}},}\ }\href {\doibase
  10.1103/PhysRevLett.116.221101} {\bibfield  {journal} {\bibinfo  {journal}
  {Phys. Rev. Lett.}\ }\textbf {\bibinfo {volume} {116}},\ \bibinfo {pages}
  {221101} (\bibinfo {year} {2016})},\ \bibinfo {note} {[Erratum:
  Phys.Rev.Lett. 121, 129902 (2018)]},\ \Eprint
  {http://arxiv.org/abs/1602.03841} {arXiv:1602.03841 [gr-qc]} \BibitemShut
  {NoStop}%
\bibitem [{\citenamefont {Abbott}\ \emph {et~al.}(2021)\citenamefont {Abbott}
  \emph {et~al.}}]{LIGOScientific:2021sio}%
  \BibitemOpen
  \bibfield  {author} {\bibinfo {author} {\bibfnamefont {R.}~\bibnamefont
  {Abbott}} \emph {et~al.} (\bibinfo {collaboration} {LIGO Scientific, VIRGO,
  KAGRA}),\ }\href@noop {} {\enquote {\bibinfo {title} {{Tests of General
  Relativity with GWTC-3}},}\ } (\bibinfo {year} {2021}),\ \Eprint
  {http://arxiv.org/abs/2112.06861} {arXiv:2112.06861 [gr-qc]} \BibitemShut
  {NoStop}%
\bibitem [{\citenamefont {{Sazhin}}(1978)}]{1978SvA....22...36S}%
  \BibitemOpen
  \bibfield  {author} {\bibinfo {author} {\bibfnamefont {M.~V.}\ \bibnamefont
  {{Sazhin}}},\ }\bibfield  {title} {\enquote {\bibinfo {title} {{Opportunities
  for detecting ultralong gravitational waves}},}\ }\href@noop {} {\bibfield
  {journal} {\bibinfo  {journal} {\sovast}\ }\textbf {\bibinfo {volume} {22}},\
  \bibinfo {pages} {36--38} (\bibinfo {year} {1978})}\BibitemShut {NoStop}%
\bibitem [{\citenamefont {Detweiler}(1979)}]{Detweiler:1979wn}%
  \BibitemOpen
  \bibfield  {author} {\bibinfo {author} {\bibfnamefont {Steven~L.}\
  \bibnamefont {Detweiler}},\ }\bibfield  {title} {\enquote {\bibinfo {title}
  {{Pulsar timing measurements and the search for gravitational waves}},}\
  }\href {\doibase 10.1086/157593} {\bibfield  {journal} {\bibinfo  {journal}
  {Astrophys. J.}\ }\textbf {\bibinfo {volume} {234}},\ \bibinfo {pages}
  {1100--1104} (\bibinfo {year} {1979})}\BibitemShut {NoStop}%
\bibitem [{\citenamefont {{Foster}}\ and\ \citenamefont
  {{Backer}}(1990)}]{1990ApJ...361..300F}%
  \BibitemOpen
  \bibfield  {author} {\bibinfo {author} {\bibfnamefont {R.~S.}\ \bibnamefont
  {{Foster}}}\ and\ \bibinfo {author} {\bibfnamefont {D.~C.}\ \bibnamefont
  {{Backer}}},\ }\bibfield  {title} {\enquote {\bibinfo {title} {{Constructing
  a Pulsar Timing Array}},}\ }\href {\doibase 10.1086/169195} {\bibfield
  {journal} {\bibinfo  {journal} {\apj}\ }\textbf {\bibinfo {volume} {361}},\
  \bibinfo {pages} {300} (\bibinfo {year} {1990})}\BibitemShut {NoStop}%
\bibitem [{\citenamefont {Bi}\ \emph {et~al.}(2023)\citenamefont {Bi},
  \citenamefont {Wu}, \citenamefont {Chen},\ and\ \citenamefont
  {Huang}}]{Bi:2023tib}%
  \BibitemOpen
  \bibfield  {author} {\bibinfo {author} {\bibfnamefont {Yan-Chen}\
  \bibnamefont {Bi}}, \bibinfo {author} {\bibfnamefont {Yu-Mei}\ \bibnamefont
  {Wu}}, \bibinfo {author} {\bibfnamefont {Zu-Cheng}\ \bibnamefont {Chen}}, \
  and\ \bibinfo {author} {\bibfnamefont {Qing-Guo}\ \bibnamefont {Huang}},\
  }\bibfield  {title} {\enquote {\bibinfo {title} {{Implications for the
  supermassive black hole binaries from the NANOGrav 15-year data set}},}\
  }\href {\doibase 10.1007/s11433-023-2252-4} {\bibfield  {journal} {\bibinfo
  {journal} {Sci. China Phys. Mech. Astron.}\ }\textbf {\bibinfo {volume}
  {66}},\ \bibinfo {pages} {120402} (\bibinfo {year} {2023})},\ \Eprint
  {http://arxiv.org/abs/2307.00722} {arXiv:2307.00722 [astro-ph.CO]}
  \BibitemShut {NoStop}%
\bibitem [{\citenamefont {Wu}\ \emph {et~al.}(2024{\natexlab{a}})\citenamefont
  {Wu}, \citenamefont {Chen},\ and\ \citenamefont {Huang}}]{Wu:2023hsa}%
  \BibitemOpen
  \bibfield  {author} {\bibinfo {author} {\bibfnamefont {Yu-Mei}\ \bibnamefont
  {Wu}}, \bibinfo {author} {\bibfnamefont {Zu-Cheng}\ \bibnamefont {Chen}}, \
  and\ \bibinfo {author} {\bibfnamefont {Qing-Guo}\ \bibnamefont {Huang}},\
  }\bibfield  {title} {\enquote {\bibinfo {title} {{Cosmological interpretation
  for the stochastic signal in pulsar timing arrays}},}\ }\href {\doibase
  10.1007/s11433-023-2298-7} {\bibfield  {journal} {\bibinfo  {journal} {Sci.
  China Phys. Mech. Astron.}\ }\textbf {\bibinfo {volume} {67}},\ \bibinfo
  {pages} {240412} (\bibinfo {year} {2024}{\natexlab{a}})},\ \Eprint
  {http://arxiv.org/abs/2307.03141} {arXiv:2307.03141 [astro-ph.CO]}
  \BibitemShut {NoStop}%
\bibitem [{\citenamefont {Taylor}\ \emph {et~al.}(2016)\citenamefont {Taylor},
  \citenamefont {Vallisneri}, \citenamefont {Ellis}, \citenamefont
  {Mingarelli}, \citenamefont {Lazio},\ and\ \citenamefont {van
  Haasteren}}]{Taylor:2015msb}%
  \BibitemOpen
  \bibfield  {author} {\bibinfo {author} {\bibfnamefont {S.~R.}\ \bibnamefont
  {Taylor}}, \bibinfo {author} {\bibfnamefont {M.}~\bibnamefont {Vallisneri}},
  \bibinfo {author} {\bibfnamefont {J.~A.}\ \bibnamefont {Ellis}}, \bibinfo
  {author} {\bibfnamefont {C.~M.~F.}\ \bibnamefont {Mingarelli}}, \bibinfo
  {author} {\bibfnamefont {T.~J.~W.}\ \bibnamefont {Lazio}}, \ and\ \bibinfo
  {author} {\bibfnamefont {R.}~\bibnamefont {van Haasteren}},\ }\bibfield
  {title} {\enquote {\bibinfo {title} {{Are we there yet? Time to detection of
  nanohertz gravitational waves based on pulsar-timing array limits}},}\ }\href
  {\doibase 10.3847/2041-8205/819/1/L6} {\bibfield  {journal} {\bibinfo
  {journal} {Astrophys. J. Lett.}\ }\textbf {\bibinfo {volume} {819}},\
  \bibinfo {pages} {L6} (\bibinfo {year} {2016})},\ \Eprint
  {http://arxiv.org/abs/1511.05564} {arXiv:1511.05564 [astro-ph.IM]}
  \BibitemShut {NoStop}%
\bibitem [{\citenamefont {Burke-Spolaor}\ \emph {et~al.}(2019)\citenamefont
  {Burke-Spolaor} \emph {et~al.}}]{Burke-Spolaor:2018bvk}%
  \BibitemOpen
  \bibfield  {author} {\bibinfo {author} {\bibfnamefont {Sarah}\ \bibnamefont
  {Burke-Spolaor}} \emph {et~al.},\ }\bibfield  {title} {\enquote {\bibinfo
  {title} {{The Astrophysics of Nanohertz Gravitational Waves}},}\ }\href
  {\doibase 10.1007/s00159-019-0115-7} {\bibfield  {journal} {\bibinfo
  {journal} {Astron. Astrophys. Rev.}\ }\textbf {\bibinfo {volume} {27}},\
  \bibinfo {pages} {5} (\bibinfo {year} {2019})},\ \Eprint
  {http://arxiv.org/abs/1811.08826} {arXiv:1811.08826 [astro-ph.HE]}
  \BibitemShut {NoStop}%
\bibitem [{\citenamefont {Agazie}\ \emph
  {et~al.}(2023{\natexlab{a}})\citenamefont {Agazie} \emph
  {et~al.}}]{NANOGrav:2023hde}%
  \BibitemOpen
  \bibfield  {author} {\bibinfo {author} {\bibfnamefont {Gabriella}\
  \bibnamefont {Agazie}} \emph {et~al.} (\bibinfo {collaboration} {NANOGrav}),\
  }\bibfield  {title} {\enquote {\bibinfo {title} {{The NANOGrav 15-year Data
  Set: Observations and Timing of 68 Millisecond Pulsars}},}\ }\href {\doibase
  10.3847/2041-8213/acda9a} {\bibfield  {journal} {\bibinfo  {journal}
  {Astrophys. J. Lett.}\ }\textbf {\bibinfo {volume} {951}} (\bibinfo {year}
  {2023}{\natexlab{a}}),\ 10.3847/2041-8213/acda9a},\ \Eprint
  {http://arxiv.org/abs/2306.16217} {arXiv:2306.16217 [astro-ph.HE]}
  \BibitemShut {NoStop}%
\bibitem [{\citenamefont {Agazie}\ \emph
  {et~al.}(2023{\natexlab{b}})\citenamefont {Agazie} \emph
  {et~al.}}]{NANOGrav:2023gor}%
  \BibitemOpen
  \bibfield  {author} {\bibinfo {author} {\bibfnamefont {Gabriella}\
  \bibnamefont {Agazie}} \emph {et~al.} (\bibinfo {collaboration} {NANOGrav}),\
  }\bibfield  {title} {\enquote {\bibinfo {title} {{The NANOGrav 15-year Data
  Set: Evidence for a Gravitational-Wave Background}},}\ }\href {\doibase
  10.3847/2041-8213/acdac6} {\bibfield  {journal} {\bibinfo  {journal}
  {Astrophys. J. Lett.}\ }\textbf {\bibinfo {volume} {951}} (\bibinfo {year}
  {2023}{\natexlab{b}}),\ 10.3847/2041-8213/acdac6},\ \Eprint
  {http://arxiv.org/abs/2306.16213} {arXiv:2306.16213 [astro-ph.HE]}
  \BibitemShut {NoStop}%
\bibitem [{\citenamefont {Antoniadis}\ \emph
  {et~al.}(2023{\natexlab{a}})\citenamefont {Antoniadis} \emph
  {et~al.}}]{EPTA:2023sfo}%
  \BibitemOpen
  \bibfield  {author} {\bibinfo {author} {\bibfnamefont {J.}~\bibnamefont
  {Antoniadis}} \emph {et~al.} (\bibinfo {collaboration} {EPTA}),\ }\bibfield
  {title} {\enquote {\bibinfo {title} {{The second data release from the
  European Pulsar Timing Array - I. The dataset and timing analysis}},}\ }\href
  {\doibase 10.1051/0004-6361/202346841} {\bibfield  {journal} {\bibinfo
  {journal} {Astron. Astrophys.}\ }\textbf {\bibinfo {volume} {678}},\ \bibinfo
  {pages} {A48} (\bibinfo {year} {2023}{\natexlab{a}})},\ \Eprint
  {http://arxiv.org/abs/2306.16224} {arXiv:2306.16224 [astro-ph.HE]}
  \BibitemShut {NoStop}%
\bibitem [{\citenamefont {Antoniadis}\ \emph
  {et~al.}(2023{\natexlab{b}})\citenamefont {Antoniadis} \emph
  {et~al.}}]{EPTA:2023fyk}%
  \BibitemOpen
  \bibfield  {author} {\bibinfo {author} {\bibfnamefont {J.}~\bibnamefont
  {Antoniadis}} \emph {et~al.} (\bibinfo {collaboration} {EPTA, InPTA:}),\
  }\bibfield  {title} {\enquote {\bibinfo {title} {{The second data release
  from the European Pulsar Timing Array - III. Search for gravitational wave
  signals}},}\ }\href {\doibase 10.1051/0004-6361/202346844} {\bibfield
  {journal} {\bibinfo  {journal} {Astron. Astrophys.}\ }\textbf {\bibinfo
  {volume} {678}},\ \bibinfo {pages} {A50} (\bibinfo {year}
  {2023}{\natexlab{b}})},\ \Eprint {http://arxiv.org/abs/2306.16214}
  {arXiv:2306.16214 [astro-ph.HE]} \BibitemShut {NoStop}%
\bibitem [{\citenamefont {Zic}\ \emph {et~al.}(2023)\citenamefont {Zic} \emph
  {et~al.}}]{Zic:2023gta}%
  \BibitemOpen
  \bibfield  {author} {\bibinfo {author} {\bibfnamefont {Andrew}\ \bibnamefont
  {Zic}} \emph {et~al.},\ }\bibfield  {title} {\enquote {\bibinfo {title} {{The
  Parkes Pulsar Timing Array third data release}},}\ }\href {\doibase
  10.1017/pasa.2023.36} {\bibfield  {journal} {\bibinfo  {journal} {Publ.
  Astron. Soc. Austral.}\ }\textbf {\bibinfo {volume} {40}},\ \bibinfo {pages}
  {e049} (\bibinfo {year} {2023})},\ \Eprint {http://arxiv.org/abs/2306.16230}
  {arXiv:2306.16230 [astro-ph.HE]} \BibitemShut {NoStop}%
\bibitem [{\citenamefont {Reardon}\ \emph {et~al.}(2023)\citenamefont {Reardon}
  \emph {et~al.}}]{Reardon:2023gzh}%
  \BibitemOpen
  \bibfield  {author} {\bibinfo {author} {\bibfnamefont {Daniel~J.}\
  \bibnamefont {Reardon}} \emph {et~al.},\ }\bibfield  {title} {\enquote
  {\bibinfo {title} {{Search for an isotropic gravitational-wave background
  with the Parkes Pulsar Timing Array}},}\ }\href {\doibase
  10.3847/2041-8213/acdd02} {\bibfield  {journal} {\bibinfo  {journal}
  {Astrophys. J. Lett.}\ }\textbf {\bibinfo {volume} {951}} (\bibinfo {year}
  {2023}),\ 10.3847/2041-8213/acdd02},\ \Eprint
  {http://arxiv.org/abs/2306.16215} {arXiv:2306.16215 [astro-ph.HE]}
  \BibitemShut {NoStop}%
\bibitem [{\citenamefont {Xu}\ \emph {et~al.}(2023)\citenamefont {Xu} \emph
  {et~al.}}]{Xu:2023wog}%
  \BibitemOpen
  \bibfield  {author} {\bibinfo {author} {\bibfnamefont {Heng}\ \bibnamefont
  {Xu}} \emph {et~al.},\ }\bibfield  {title} {\enquote {\bibinfo {title}
  {{Searching for the Nano-Hertz Stochastic Gravitational Wave Background with
  the Chinese Pulsar Timing Array Data Release I}},}\ }\href {\doibase
  10.1088/1674-4527/acdfa5} {\bibfield  {journal} {\bibinfo  {journal} {Res.
  Astron. Astrophys.}\ }\textbf {\bibinfo {volume} {23}},\ \bibinfo {pages}
  {075024} (\bibinfo {year} {2023})},\ \Eprint
  {http://arxiv.org/abs/2306.16216} {arXiv:2306.16216 [astro-ph.HE]}
  \BibitemShut {NoStop}%
\bibitem [{\citenamefont {Hellings}\ and\ \citenamefont
  {Downs}(1983)}]{Hellings:1983fr}%
  \BibitemOpen
  \bibfield  {author} {\bibinfo {author} {\bibfnamefont {R.~w.}\ \bibnamefont
  {Hellings}}\ and\ \bibinfo {author} {\bibfnamefont {G.~s.}\ \bibnamefont
  {Downs}},\ }\bibfield  {title} {\enquote {\bibinfo {title} {{UPPER LIMITS ON
  THE ISOTROPIC GRAVITATIONAL RADIATION BACKGROUND FROM PULSAR TIMING
  ANALYSIS}},}\ }\href {\doibase 10.1086/183954} {\bibfield  {journal}
  {\bibinfo  {journal} {Astrophys. J.}\ }\textbf {\bibinfo {volume} {265}},\
  \bibinfo {pages} {L39--L42} (\bibinfo {year} {1983})}\BibitemShut {NoStop}%
\bibitem [{\citenamefont {Bi}\ \emph {et~al.}(2024)\citenamefont {Bi},
  \citenamefont {Wu}, \citenamefont {Chen},\ and\ \citenamefont
  {Huang}}]{Bi:2023ewq}%
  \BibitemOpen
  \bibfield  {author} {\bibinfo {author} {\bibfnamefont {Yan-Chen}\
  \bibnamefont {Bi}}, \bibinfo {author} {\bibfnamefont {Yu-Mei}\ \bibnamefont
  {Wu}}, \bibinfo {author} {\bibfnamefont {Zu-Cheng}\ \bibnamefont {Chen}}, \
  and\ \bibinfo {author} {\bibfnamefont {Qing-Guo}\ \bibnamefont {Huang}},\
  }\bibfield  {title} {\enquote {\bibinfo {title} {{Constraints on the velocity
  of gravitational waves from the NANOGrav 15-year data set}},}\ }\href
  {\doibase 10.1103/PhysRevD.109.L061101} {\bibfield  {journal} {\bibinfo
  {journal} {Phys. Rev. D}\ }\textbf {\bibinfo {volume} {109}},\ \bibinfo
  {pages} {L061101} (\bibinfo {year} {2024})},\ \Eprint
  {http://arxiv.org/abs/2310.08366} {arXiv:2310.08366 [astro-ph.CO]}
  \BibitemShut {NoStop}%
\bibitem [{\citenamefont {Chen}\ \emph {et~al.}(2024)\citenamefont {Chen},
  \citenamefont {Wu}, \citenamefont {Bi},\ and\ \citenamefont
  {Huang}}]{Chen:2023uiz}%
  \BibitemOpen
  \bibfield  {author} {\bibinfo {author} {\bibfnamefont {Zu-Cheng}\
  \bibnamefont {Chen}}, \bibinfo {author} {\bibfnamefont {Yu-Mei}\ \bibnamefont
  {Wu}}, \bibinfo {author} {\bibfnamefont {Yan-Chen}\ \bibnamefont {Bi}}, \
  and\ \bibinfo {author} {\bibfnamefont {Qing-Guo}\ \bibnamefont {Huang}},\
  }\bibfield  {title} {\enquote {\bibinfo {title} {{Search for nontensorial
  gravitational-wave backgrounds in the NANOGrav 15-year dataset}},}\ }\href
  {\doibase 10.1103/PhysRevD.109.084045} {\bibfield  {journal} {\bibinfo
  {journal} {Phys. Rev. D}\ }\textbf {\bibinfo {volume} {109}},\ \bibinfo
  {pages} {084045} (\bibinfo {year} {2024})},\ \Eprint
  {http://arxiv.org/abs/2310.11238} {arXiv:2310.11238 [astro-ph.CO]}
  \BibitemShut {NoStop}%
\bibitem [{\citenamefont {Wu}\ \emph {et~al.}(2022)\citenamefont {Wu},
  \citenamefont {Chen},\ and\ \citenamefont {Huang}}]{Wu:2021kmd}%
  \BibitemOpen
  \bibfield  {author} {\bibinfo {author} {\bibfnamefont {Yu-Mei}\ \bibnamefont
  {Wu}}, \bibinfo {author} {\bibfnamefont {Zu-Cheng}\ \bibnamefont {Chen}}, \
  and\ \bibinfo {author} {\bibfnamefont {Qing-Guo}\ \bibnamefont {Huang}},\
  }\bibfield  {title} {\enquote {\bibinfo {title} {{Constraining the
  Polarization of Gravitational Waves with the Parkes Pulsar Timing Array
  Second Data Release}},}\ }\href {\doibase 10.3847/1538-4357/ac35cc}
  {\bibfield  {journal} {\bibinfo  {journal} {Astrophys. J.}\ }\textbf
  {\bibinfo {volume} {925}},\ \bibinfo {pages} {37} (\bibinfo {year} {2022})},\
  \Eprint {http://arxiv.org/abs/2108.10518} {arXiv:2108.10518 [astro-ph.CO]}
  \BibitemShut {NoStop}%
\bibitem [{\citenamefont {Chen}\ \emph {et~al.}(2022)\citenamefont {Chen},
  \citenamefont {Wu},\ and\ \citenamefont {Huang}}]{Chen:2021ncc}%
  \BibitemOpen
  \bibfield  {author} {\bibinfo {author} {\bibfnamefont {Zu-Cheng}\
  \bibnamefont {Chen}}, \bibinfo {author} {\bibfnamefont {Yu-Mei}\ \bibnamefont
  {Wu}}, \ and\ \bibinfo {author} {\bibfnamefont {Qing-Guo}\ \bibnamefont
  {Huang}},\ }\bibfield  {title} {\enquote {\bibinfo {title} {{Searching for
  isotropic stochastic gravitational-wave background in the international
  pulsar timing array second data release}},}\ }\href {\doibase
  10.1088/1572-9494/ac7cdf} {\bibfield  {journal} {\bibinfo  {journal} {Commun.
  Theor. Phys.}\ }\textbf {\bibinfo {volume} {74}},\ \bibinfo {pages} {105402}
  (\bibinfo {year} {2022})},\ \Eprint {http://arxiv.org/abs/2109.00296}
  {arXiv:2109.00296 [astro-ph.CO]} \BibitemShut {NoStop}%
\bibitem [{\citenamefont {Lee}\ \emph {et~al.}(2010)\citenamefont {Lee},
  \citenamefont {Jenet}, \citenamefont {Price}, \citenamefont {Wex},\ and\
  \citenamefont {Kramer}}]{Lee:2010cg}%
  \BibitemOpen
  \bibfield  {author} {\bibinfo {author} {\bibfnamefont {Kejia}\ \bibnamefont
  {Lee}}, \bibinfo {author} {\bibfnamefont {Fredrick~A.}\ \bibnamefont
  {Jenet}}, \bibinfo {author} {\bibfnamefont {Richard~H.}\ \bibnamefont
  {Price}}, \bibinfo {author} {\bibfnamefont {Norbert}\ \bibnamefont {Wex}}, \
  and\ \bibinfo {author} {\bibfnamefont {Michael}\ \bibnamefont {Kramer}},\
  }\bibfield  {title} {\enquote {\bibinfo {title} {{Detecting massive gravitons
  using pulsar timing arrays}},}\ }\href {\doibase
  10.1088/0004-637X/722/2/1589} {\bibfield  {journal} {\bibinfo  {journal}
  {Astrophys. J.}\ }\textbf {\bibinfo {volume} {722}},\ \bibinfo {pages}
  {1589--1597} (\bibinfo {year} {2010})},\ \Eprint
  {http://arxiv.org/abs/1008.2561} {arXiv:1008.2561 [astro-ph.HE]} \BibitemShut
  {NoStop}%
\bibitem [{\citenamefont {Liang}\ and\ \citenamefont
  {Trodden}(2021)}]{Liang:2021bct}%
  \BibitemOpen
  \bibfield  {author} {\bibinfo {author} {\bibfnamefont {Qiuyue}\ \bibnamefont
  {Liang}}\ and\ \bibinfo {author} {\bibfnamefont {Mark}\ \bibnamefont
  {Trodden}},\ }\bibfield  {title} {\enquote {\bibinfo {title} {{Detecting the
  stochastic gravitational wave background from massive gravity with pulsar
  timing arrays}},}\ }\href {\doibase 10.1103/PhysRevD.104.084052} {\bibfield
  {journal} {\bibinfo  {journal} {Phys. Rev. D}\ }\textbf {\bibinfo {volume}
  {104}},\ \bibinfo {pages} {084052} (\bibinfo {year} {2021})},\ \Eprint
  {http://arxiv.org/abs/2108.05344} {arXiv:2108.05344 [astro-ph.CO]}
  \BibitemShut {NoStop}%
\bibitem [{\citenamefont {Wu}\ \emph {et~al.}(2023)\citenamefont {Wu},
  \citenamefont {Chen},\ and\ \citenamefont {Huang}}]{Wu:2023pbt}%
  \BibitemOpen
  \bibfield  {author} {\bibinfo {author} {\bibfnamefont {Yu-Mei}\ \bibnamefont
  {Wu}}, \bibinfo {author} {\bibfnamefont {Zu-Cheng}\ \bibnamefont {Chen}}, \
  and\ \bibinfo {author} {\bibfnamefont {Qing-Guo}\ \bibnamefont {Huang}},\
  }\bibfield  {title} {\enquote {\bibinfo {title} {{Search for stochastic
  gravitational-wave background from massive gravity in the NANOGrav 12.5-year
  dataset}},}\ }\href {\doibase 10.1103/PhysRevD.107.042003} {\bibfield
  {journal} {\bibinfo  {journal} {Phys. Rev. D}\ }\textbf {\bibinfo {volume}
  {107}},\ \bibinfo {pages} {042003} (\bibinfo {year} {2023})},\ \Eprint
  {http://arxiv.org/abs/2302.00229} {arXiv:2302.00229 [gr-qc]} \BibitemShut
  {NoStop}%
\bibitem [{\citenamefont {Wu}\ \emph {et~al.}(2024{\natexlab{b}})\citenamefont
  {Wu}, \citenamefont {Chen}, \citenamefont {Bi},\ and\ \citenamefont
  {Huang}}]{Wu:2023rib}%
  \BibitemOpen
  \bibfield  {author} {\bibinfo {author} {\bibfnamefont {Yu-Mei}\ \bibnamefont
  {Wu}}, \bibinfo {author} {\bibfnamefont {Zu-Cheng}\ \bibnamefont {Chen}},
  \bibinfo {author} {\bibfnamefont {Yan-Chen}\ \bibnamefont {Bi}}, \ and\
  \bibinfo {author} {\bibfnamefont {Qing-Guo}\ \bibnamefont {Huang}},\
  }\bibfield  {title} {\enquote {\bibinfo {title} {{Constraining the graviton
  mass with the NANOGrav 15\,year data set}},}\ }\href {\doibase
  10.1088/1361-6382/ad2a9b} {\bibfield  {journal} {\bibinfo  {journal} {Class.
  Quant. Grav.}\ }\textbf {\bibinfo {volume} {41}},\ \bibinfo {pages} {075002}
  (\bibinfo {year} {2024}{\natexlab{b}})},\ \Eprint
  {http://arxiv.org/abs/2310.07469} {arXiv:2310.07469 [astro-ph.CO]}
  \BibitemShut {NoStop}%
\bibitem [{\citenamefont {Roebber}\ and\ \citenamefont
  {Holder}(2017)}]{Roebber:2016jzl}%
  \BibitemOpen
  \bibfield  {author} {\bibinfo {author} {\bibfnamefont {Elinore}\ \bibnamefont
  {Roebber}}\ and\ \bibinfo {author} {\bibfnamefont {Gilbert}\ \bibnamefont
  {Holder}},\ }\bibfield  {title} {\enquote {\bibinfo {title} {{Harmonic space
  analysis of pulsar timing array redshift maps}},}\ }\href {\doibase
  10.3847/1538-4357/835/1/21} {\bibfield  {journal} {\bibinfo  {journal}
  {Astrophys. J.}\ }\textbf {\bibinfo {volume} {835}},\ \bibinfo {pages} {21}
  (\bibinfo {year} {2017})},\ \Eprint {http://arxiv.org/abs/1609.06758}
  {arXiv:1609.06758 [astro-ph.CO]} \BibitemShut {NoStop}%
\bibitem [{\citenamefont {Allen}(2023)}]{Allen:2022dzg}%
  \BibitemOpen
  \bibfield  {author} {\bibinfo {author} {\bibfnamefont {Bruce}\ \bibnamefont
  {Allen}},\ }\bibfield  {title} {\enquote {\bibinfo {title} {{Variance of the
  Hellings-Downs correlation}},}\ }\href {\doibase 10.1103/PhysRevD.107.043018}
  {\bibfield  {journal} {\bibinfo  {journal} {Phys. Rev. D}\ }\textbf {\bibinfo
  {volume} {107}},\ \bibinfo {pages} {043018} (\bibinfo {year} {2023})},\
  \Eprint {http://arxiv.org/abs/2205.05637} {arXiv:2205.05637 [gr-qc]}
  \BibitemShut {NoStop}%
\bibitem [{\citenamefont {Romano}\ and\ \citenamefont
  {Allen}(2024)}]{Romano:2023zhb}%
  \BibitemOpen
  \bibfield  {author} {\bibinfo {author} {\bibfnamefont {Joseph~D.}\
  \bibnamefont {Romano}}\ and\ \bibinfo {author} {\bibfnamefont {Bruce}\
  \bibnamefont {Allen}},\ }\bibfield  {title} {\enquote {\bibinfo {title}
  {{Answers to frequently asked questions about the pulsar timing array
  Hellings and Downs curve}},}\ }\href {\doibase 10.1088/1361-6382/ad4c4c}
  {\bibfield  {journal} {\bibinfo  {journal} {Class. Quant. Grav.}\ }\textbf
  {\bibinfo {volume} {41}},\ \bibinfo {pages} {175008} (\bibinfo {year}
  {2024})},\ \Eprint {http://arxiv.org/abs/2308.05847} {arXiv:2308.05847
  [gr-qc]} \BibitemShut {NoStop}%
\bibitem [{\citenamefont {Bernardo}\ and\ \citenamefont
  {Ng}(2022)}]{Bernardo:2022xzl}%
  \BibitemOpen
  \bibfield  {author} {\bibinfo {author} {\bibfnamefont {Reginald~Christian}\
  \bibnamefont {Bernardo}}\ and\ \bibinfo {author} {\bibfnamefont {Kin-Wang}\
  \bibnamefont {Ng}},\ }\bibfield  {title} {\enquote {\bibinfo {title} {{Pulsar
  and cosmic variances of pulsar timing-array correlation measurements of the
  stochastic gravitational wave background}},}\ }\href {\doibase
  10.1088/1475-7516/2022/11/046} {\bibfield  {journal} {\bibinfo  {journal}
  {JCAP}\ }\textbf {\bibinfo {volume} {11}},\ \bibinfo {pages} {046} (\bibinfo
  {year} {2022})},\ \Eprint {http://arxiv.org/abs/2209.14834} {arXiv:2209.14834
  [gr-qc]} \BibitemShut {NoStop}%
\bibitem [{\citenamefont {Bernardo}\ and\ \citenamefont
  {Ng}(2023{\natexlab{a}})}]{Bernardo:2023bqx}%
  \BibitemOpen
  \bibfield  {author} {\bibinfo {author} {\bibfnamefont {Reginald~Christian}\
  \bibnamefont {Bernardo}}\ and\ \bibinfo {author} {\bibfnamefont {Kin-Wang}\
  \bibnamefont {Ng}},\ }\bibfield  {title} {\enquote {\bibinfo {title}
  {{Hunting the stochastic gravitational wave background in pulsar timing array
  cross correlations through theoretical uncertainty}},}\ }\href {\doibase
  10.1088/1475-7516/2023/08/028} {\bibfield  {journal} {\bibinfo  {journal}
  {JCAP}\ }\textbf {\bibinfo {volume} {08}},\ \bibinfo {pages} {028} (\bibinfo
  {year} {2023}{\natexlab{a}})},\ \Eprint {http://arxiv.org/abs/2304.07040}
  {arXiv:2304.07040 [gr-qc]} \BibitemShut {NoStop}%
\bibitem [{\citenamefont {Wu}\ \emph {et~al.}(2024{\natexlab{c}})\citenamefont
  {Wu}, \citenamefont {Bi},\ and\ \citenamefont {Huang}}]{Wu:2024xkp}%
  \BibitemOpen
  \bibfield  {author} {\bibinfo {author} {\bibfnamefont {Yu-Mei}\ \bibnamefont
  {Wu}}, \bibinfo {author} {\bibfnamefont {Yan-Chen}\ \bibnamefont {Bi}}, \
  and\ \bibinfo {author} {\bibfnamefont {Qing-Guo}\ \bibnamefont {Huang}},\
  }\href@noop {} {\enquote {\bibinfo {title} {{The spatial correlations between
  pulsars for interfering sources in Pulsar Timing Array and evidence for
  gravitational-wave background in NANOGrav 15-year data set}},}\ } (\bibinfo
  {year} {2024}{\natexlab{c}}),\ \Eprint {http://arxiv.org/abs/2407.07319}
  {arXiv:2407.07319 [astro-ph.CO]} \BibitemShut {NoStop}%
\bibitem [{\citenamefont {Bernardo}\ and\ \citenamefont
  {Ng}(2024{\natexlab{a}})}]{Bernardo:2023pwt}%
  \BibitemOpen
  \bibfield  {author} {\bibinfo {author} {\bibfnamefont {Reginald~Christian}\
  \bibnamefont {Bernardo}}\ and\ \bibinfo {author} {\bibfnamefont {Kin-Wang}\
  \bibnamefont {Ng}},\ }\bibfield  {title} {\enquote {\bibinfo {title}
  {{Testing gravity with cosmic variance-limited pulsar timing array
  correlations}},}\ }\href {\doibase 10.1103/PhysRevD.109.L101502} {\bibfield
  {journal} {\bibinfo  {journal} {Phys. Rev. D}\ }\textbf {\bibinfo {volume}
  {109}},\ \bibinfo {pages} {L101502} (\bibinfo {year} {2024}{\natexlab{a}})},\
  \Eprint {http://arxiv.org/abs/2306.13593} {arXiv:2306.13593 [gr-qc]}
  \BibitemShut {NoStop}%
\bibitem [{\citenamefont {Bernardo}\ and\ \citenamefont
  {Ng}(2024{\natexlab{b}})}]{Bernardo:2023zna}%
  \BibitemOpen
  \bibfield  {author} {\bibinfo {author} {\bibfnamefont {Reginald~Christian}\
  \bibnamefont {Bernardo}}\ and\ \bibinfo {author} {\bibfnamefont {Kin-Wang}\
  \bibnamefont {Ng}},\ }\bibfield  {title} {\enquote {\bibinfo {title} {{Beyond
  the Hellings\textendash{}Downs curve: Non-Einsteinian gravitational waves in
  pulsar timing array correlations}},}\ }\href {\doibase
  10.1051/0004-6361/202449483} {\bibfield  {journal} {\bibinfo  {journal}
  {Astron. Astrophys.}\ }\textbf {\bibinfo {volume} {691}},\ \bibinfo {pages}
  {A126} (\bibinfo {year} {2024}{\natexlab{b}})},\ \Eprint
  {http://arxiv.org/abs/2310.07537} {arXiv:2310.07537 [gr-qc]} \BibitemShut
  {NoStop}%
\bibitem [{\citenamefont {Bernardo}\ \emph {et~al.}(2024)\citenamefont
  {Bernardo}, \citenamefont {Liu},\ and\ \citenamefont
  {Ng}}]{Bernardo:2023jhs}%
  \BibitemOpen
  \bibfield  {author} {\bibinfo {author} {\bibfnamefont {Reginald~Christian}\
  \bibnamefont {Bernardo}}, \bibinfo {author} {\bibfnamefont {Guo-Chin}\
  \bibnamefont {Liu}}, \ and\ \bibinfo {author} {\bibfnamefont {Kin-Wang}\
  \bibnamefont {Ng}},\ }\bibfield  {title} {\enquote {\bibinfo {title}
  {{Correlations for an anisotropic polarized stochastic gravitational wave
  background in pulsar timing arrays}},}\ }\href {\doibase
  10.1088/1475-7516/2024/04/034} {\bibfield  {journal} {\bibinfo  {journal}
  {JCAP}\ }\textbf {\bibinfo {volume} {04}},\ \bibinfo {pages} {034} (\bibinfo
  {year} {2024})},\ \Eprint {http://arxiv.org/abs/2312.03383} {arXiv:2312.03383
  [gr-qc]} \BibitemShut {NoStop}%
\bibitem [{\citenamefont {Bernardo}\ and\ \citenamefont
  {Ng}(2025)}]{Bernardo:2024bdc}%
  \BibitemOpen
  \bibfield  {author} {\bibinfo {author} {\bibfnamefont {Reginald~Christian}\
  \bibnamefont {Bernardo}}\ and\ \bibinfo {author} {\bibfnamefont {Kin-Wang}\
  \bibnamefont {Ng}},\ }\bibfield  {title} {\enquote {\bibinfo {title}
  {{Charting the nanohertz gravitational wave sky with pulsar timing
  arrays}},}\ }\href {\doibase 10.1142/S0218271825400139} {\bibfield  {journal}
  {\bibinfo  {journal} {Int. J. Mod. Phys. D}\ }\textbf {\bibinfo {volume}
  {34}},\ \bibinfo {pages} {2540013} (\bibinfo {year} {2025})},\ \Eprint
  {http://arxiv.org/abs/2409.07955} {arXiv:2409.07955 [astro-ph.CO]}
  \BibitemShut {NoStop}%
\bibitem [{\citenamefont {de~Rham}(2014)}]{deRham:2014zqa}%
  \BibitemOpen
  \bibfield  {author} {\bibinfo {author} {\bibfnamefont {Claudia}\ \bibnamefont
  {de~Rham}},\ }\bibfield  {title} {\enquote {\bibinfo {title} {{Massive
  Gravity}},}\ }\href {\doibase 10.12942/lrr-2014-7} {\bibfield  {journal}
  {\bibinfo  {journal} {Living Rev. Rel.}\ }\textbf {\bibinfo {volume} {17}},\
  \bibinfo {pages} {7} (\bibinfo {year} {2014})},\ \Eprint
  {http://arxiv.org/abs/1401.4173} {arXiv:1401.4173 [hep-th]} \BibitemShut
  {NoStop}%
\bibitem [{\citenamefont {Bernardo}\ and\ \citenamefont
  {Ng}(2023{\natexlab{b}})}]{Bernardo:2023mxc}%
  \BibitemOpen
  \bibfield  {author} {\bibinfo {author} {\bibfnamefont {Reginald~Christian}\
  \bibnamefont {Bernardo}}\ and\ \bibinfo {author} {\bibfnamefont {Kin-Wang}\
  \bibnamefont {Ng}},\ }\bibfield  {title} {\enquote {\bibinfo {title}
  {{Constraining gravitational wave propagation using pulsar timing array
  correlations}},}\ }\href {\doibase 10.1103/PhysRevD.107.L101502} {\bibfield
  {journal} {\bibinfo  {journal} {Phys. Rev. D}\ }\textbf {\bibinfo {volume}
  {107}},\ \bibinfo {pages} {L101502} (\bibinfo {year} {2023}{\natexlab{b}})},\
  \Eprint {http://arxiv.org/abs/2302.11796} {arXiv:2302.11796 [gr-qc]}
  \BibitemShut {NoStop}%
\bibitem [{\citenamefont {Chamberlin}\ and\ \citenamefont
  {Siemens}(2012)}]{Chamberlin:2011ev}%
  \BibitemOpen
  \bibfield  {author} {\bibinfo {author} {\bibfnamefont {Sydney~J.}\
  \bibnamefont {Chamberlin}}\ and\ \bibinfo {author} {\bibfnamefont {Xavier}\
  \bibnamefont {Siemens}},\ }\bibfield  {title} {\enquote {\bibinfo {title}
  {{Stochastic backgrounds in alternative theories of gravity: overlap
  reduction functions for pulsar timing arrays}},}\ }\href {\doibase
  10.1103/PhysRevD.85.082001} {\bibfield  {journal} {\bibinfo  {journal} {Phys.
  Rev. D}\ }\textbf {\bibinfo {volume} {85}},\ \bibinfo {pages} {082001}
  (\bibinfo {year} {2012})},\ \Eprint {http://arxiv.org/abs/1111.5661}
  {arXiv:1111.5661 [astro-ph.HE]} \BibitemShut {NoStop}%
\bibitem [{\citenamefont {Cornish}\ and\ \citenamefont
  {Sesana}(2013)}]{Cornish:2013aba}%
  \BibitemOpen
  \bibfield  {author} {\bibinfo {author} {\bibfnamefont {Neil~J.}\ \bibnamefont
  {Cornish}}\ and\ \bibinfo {author} {\bibfnamefont {A.}~\bibnamefont
  {Sesana}},\ }\bibfield  {title} {\enquote {\bibinfo {title} {{Pulsar Timing
  Array Analysis for Black Hole Backgrounds}},}\ }\href {\doibase
  10.1088/0264-9381/30/22/224005} {\bibfield  {journal} {\bibinfo  {journal}
  {Class. Quant. Grav.}\ }\textbf {\bibinfo {volume} {30}},\ \bibinfo {pages}
  {224005} (\bibinfo {year} {2013})},\ \Eprint {http://arxiv.org/abs/1305.0326}
  {arXiv:1305.0326 [gr-qc]} \BibitemShut {NoStop}%
\bibitem [{\citenamefont {Bernardo}\ and\ \citenamefont
  {Ng}(2023{\natexlab{c}})}]{Bernardo:2022rif}%
  \BibitemOpen
  \bibfield  {author} {\bibinfo {author} {\bibfnamefont {Reginald~Christian}\
  \bibnamefont {Bernardo}}\ and\ \bibinfo {author} {\bibfnamefont {Kin-Wang}\
  \bibnamefont {Ng}},\ }\bibfield  {title} {\enquote {\bibinfo {title}
  {{Stochastic gravitational wave background phenomenology in a pulsar timing
  array}},}\ }\href {\doibase 10.1103/PhysRevD.107.044007} {\bibfield
  {journal} {\bibinfo  {journal} {Phys. Rev. D}\ }\textbf {\bibinfo {volume}
  {107}},\ \bibinfo {pages} {044007} (\bibinfo {year} {2023}{\natexlab{c}})},\
  \Eprint {http://arxiv.org/abs/2208.12538} {arXiv:2208.12538 [gr-qc]}
  \BibitemShut {NoStop}%
\bibitem [{\citenamefont {Dom{\`e}nech}\ and\ \citenamefont
  {Tsabodimos}(2024)}]{Domenech:2024pow}%
  \BibitemOpen
  \bibfield  {author} {\bibinfo {author} {\bibfnamefont {Guillem}\ \bibnamefont
  {Dom{\`e}nech}}\ and\ \bibinfo {author} {\bibfnamefont {Apostolos}\
  \bibnamefont {Tsabodimos}},\ }\bibfield  {title} {\enquote {\bibinfo {title}
  {{Finite distance effects on the Hellings{\textendash}Downs curve in modified
  gravity}},}\ }\href {\doibase 10.1140/epjc/s10052-024-13418-w} {\bibfield
  {journal} {\bibinfo  {journal} {Eur. Phys. J. C}\ }\textbf {\bibinfo {volume}
  {84}},\ \bibinfo {pages} {1005} (\bibinfo {year} {2024})},\ \bibinfo {note}
  {[Erratum: Eur.Phys.J.C 84, 1123 (2024)]},\ \Eprint
  {http://arxiv.org/abs/2407.21567} {arXiv:2407.21567 [gr-qc]} \BibitemShut
  {NoStop}%
\bibitem [{\citenamefont {Allen}\ and\ \citenamefont
  {Romano}(2025)}]{Allen:2024uqs}%
  \BibitemOpen
  \bibfield  {author} {\bibinfo {author} {\bibfnamefont {Bruce}\ \bibnamefont
  {Allen}}\ and\ \bibinfo {author} {\bibfnamefont {Joseph~D.}\ \bibnamefont
  {Romano}},\ }\bibfield  {title} {\enquote {\bibinfo {title} {{Optimal
  Reconstruction of the Hellings and Downs Correlation}},}\ }\href {\doibase
  10.1103/PhysRevLett.134.031401} {\bibfield  {journal} {\bibinfo  {journal}
  {Phys. Rev. Lett.}\ }\textbf {\bibinfo {volume} {134}},\ \bibinfo {pages}
  {031401} (\bibinfo {year} {2025})},\ \Eprint
  {http://arxiv.org/abs/2407.10968} {arXiv:2407.10968 [gr-qc]} \BibitemShut
  {NoStop}%
\bibitem [{\citenamefont {Caprini}\ and\ \citenamefont
  {Figueroa}(2018)}]{Caprini:2018mtu}%
  \BibitemOpen
  \bibfield  {author} {\bibinfo {author} {\bibfnamefont {Chiara}\ \bibnamefont
  {Caprini}}\ and\ \bibinfo {author} {\bibfnamefont {Daniel~G.}\ \bibnamefont
  {Figueroa}},\ }\bibfield  {title} {\enquote {\bibinfo {title} {{Cosmological
  Backgrounds of Gravitational Waves}},}\ }\href {\doibase
  10.1088/1361-6382/aac608} {\bibfield  {journal} {\bibinfo  {journal} {Class.
  Quant. Grav.}\ }\textbf {\bibinfo {volume} {35}},\ \bibinfo {pages} {163001}
  (\bibinfo {year} {2018})},\ \Eprint {http://arxiv.org/abs/1801.04268}
  {arXiv:1801.04268 [astro-ph.CO]} \BibitemShut {NoStop}%
\bibitem [{\citenamefont {Wu}(2025)}]{massive_data}%
  \BibitemOpen
  \bibfield  {author} {\bibinfo {author} {\bibfnamefont {Yu-Mei}\ \bibnamefont
  {Wu}},\ }\href
  {https://github.com/wym-7006/Interference_Effect_in_Massive_Gravity}
  {\enquote {\bibinfo {title}
  {{Interference$\_$Effect$\_$in$\_$Massive$\_$Gravity}},}\ }\bibinfo
  {howpublished} {github} (\bibinfo {year} {2025})\BibitemShut {NoStop}%
\end{thebibliography}%
	
\end{document}